\documentclass[10pt,prb,twocolumn,notitlepage,showpacs,superscriptaddress]{revtex4-1}
\usepackage{amsthm,amsmath,amsfonts,amssymb,verbatim, color}
\usepackage{graphicx}
\usepackage{epsfig,slashed}
\usepackage[colorlinks=true,citecolor=blue,
linkcolor=blue,urlcolor=blue]{hyperref}
\begin{document}

\newcommand{\Tr}{\mathop{\mathrm{Tr}}}
\newcommand{\bsigma}{\boldsymbol{\sigma}}
\renewcommand{\Re}{\mathop{\mathrm{Re}}}
\renewcommand{\Im}{\mathop{\mathrm{Im}}}
\renewcommand{\b}[1]{\mathbf{#1}}
\newcommand{\diag}{\mathrm{diag}}
\newcommand{\sign}{\mathrm{sign}}
\newcommand{\sgn}{\mathop{\mathrm{sgn}}}

\title{Weyl semimetals with short-range interactions}

\author{Joseph Maciejko}
\affiliation{Princeton Center for Theoretical Science, Princeton University, Princeton, New Jersey 08544, USA}
\affiliation{Department of Physics, University of Alberta, Edmonton, Alberta T6G 2E1, Canada}

\author{Rahul Nandkishore}
\affiliation{Princeton Center for Theoretical Science, Princeton University, Princeton, New Jersey 08544, USA}

\date\today

\begin{abstract}
We construct a low-energy effective field theory of fermions interacting via short-range interactions in a simple two-band model of a Weyl semimetal on the cubic lattice and investigate possible broken-symmetry ground states through a one-loop renormalization group (RG) analysis. Using the symmetries of the noninteracting Hamiltonian to constrain the form of the interaction term leads to four independent coupling constants. We investigate the stability of RG flows towards strong coupling and find a single stable trajectory. In order to explore possible broken-symmetry ground states, we calculate susceptibilities in the particle-hole and particle-particle channels along this trajectory and find that the leading instability is towards a fully gapped spin-density wave (SDW) ground state. The sliding mode of this SDW couples to the external electromagnetic fields like the Peccei-Quinn axion field of particle physics. We also study the maximally symmetric version of our model with a single independent coupling constant.  Possible ground states in this case are either gapless ferromagnetic states where the spin waves couple to the Weyl fermions like the spatial components of a (possibly chiral) gauge field, or a fully gapped spin-singlet Fulde-Ferrell-Larkin-Ovchinnikov (FFLO) superconducting state.
\end{abstract}


\maketitle

\section{Introduction}
\label{sec:intro}

The consideration of topological aspects of the electronic structure of solids has led to spectacular recent developments in quantum condensed matter physics. In this context, topology refers to the invariance of certain global properties of the electronic structure under perturbations of the system that are sufficiently small and may have to preserve certain symmetries, but are otherwise arbitrary. The prime example of this is the integer quantum Hall effect,\cite{klitzing1980} where the quantization of the Hall conductance, a global property of the band structure,\cite{thouless1982} is insensitive to arbitrary perturbations. A more recent example is the discovery of topological insulators in two and three dimensions,\cite{hasan2010,qi2011} where the quantization of a $\mathbb{Z}_2$ topological invariant and associated electromagnetic response properties is insentitive to perturbations that preserve time-reversal symmetry. The quantization of a topological invariant of the bulk band structure often implies the appearance of robust gapless states on the boundary of the system that could be used for the nearly dissipationless transport of information.\cite{buttiker2009}

Given that integer quantum Hall systems and topological insulators are both band insulators, one might wonder whether such topological phenomena are limited to gapped systems. The answer is no. The stability of the Fermi surface of a metal against perturbations that preserve translation symmetry\cite{NoteSC} can be described by topological invariants.\cite{Volovik,horava2005} A semimetal, where the Fermi surface (in dimensions higher than one) reduces to a discrete set of points, is an interesting case intermediate between a metal and an insulator. In three dimensions, a linear crossing of two nondegenerate bands is stable against arbitrary weak translation symmetry preserving perturbations (for a discussion of the effects of translation symmetry breaking perturbations, see Ref.~\onlinecite{NHS2013} and references contained therein). Near the crossing point $\b{k}=\b{k}_0$, the effective Hamiltonian for these two bands is of the form\cite{wan2011}
\begin{align}\label{hweyl}
h(\b{k})=E_0+\b{v}_0\cdot(\b{k}-\b{k}_0)+
\sum_{i=1}^3\b{v}_i\cdot(\b{k}-\b{k}_0)\sigma_i,
\end{align}
where $\b{v}_0,\ldots,\b{v}_3$ are real vectors and $\sigma_1,\sigma_2,\sigma_3$ are the three Pauli matrices. Because there are as many momentum directions in 3D as there are independent, anticommuting Hermitian $2\times 2$ matrices, it is impossible to add a (mass) term to Eq.~(\ref{hweyl}) that would introduce a gap between the two bands. The Hamiltonian (\ref{hweyl}) describes a single chiral or Weyl fermion. Weyl fermions are either right-handed or left-handed, where the handedness or chirality defined by $c=\sgn(\b{v}_1\cdot(\b{v}_2\times\b{v}_3))=\pm 1$ is a topological invariant.\cite{wan2011,Volovik} A known example of 3D Weyl fermion in nature  is the nodal Bogoliubov quasiparticle in the A-phase of superfluid $^3$He.\cite{Volovik}

Recent theoretical work suggests that electronic structures with Weyl points of the type (\ref{hweyl}) occurring at the Fermi level may be realized in solid-state systems. These Weyl semimetals\cite{balents2011,vafek2013} have been predicted to occur as an intermediate gapless phase between a trivial and a topological insulator,\cite{murakami2007,singh2012} in topological insulator multilayers,\cite{burkov2011,burkov2011b,zyuzin2012b,halasz2012,lin2013} and in magnetically doped topological insulators;\cite{cho2011,liu2013,bulmash2013} in the phase diagram of pyrochlore iridates;\cite{wan2011,witczak-krempa2012,go2012,chen2012} in the ferromagnetic compounds HgCr$_2$Se$_4$\cite{xu2011} and CdO/EuO;\cite{zhang2013} and by applying a magnetic field\cite{gorbar2013} to a Dirac semimetal,\cite{young2012,wang2012,wang2013b,steinberg2013,
neupane2013,borisenko2013,liu2013b,orlita2013} where two Weyl points coexisting at the same momentum are protected by crystallographic symmetries. Recent magnetoresistance studies in Bi$_{1-x}$Sb$_x$,\cite{kim2013} as well as in the pyrochlore iridate Bi$_2$Ir$_2$O$_7$,\cite{chu2013} report observations that are consistent with the phenomenology of Weyl semimetals. For a solid-state system on a lattice, the Nielsen-Ninomiya theorem\cite{nielsen1981,nielsen1981b} implies that Weyl points must appear in pairs. Such a system with an even number of Weyl points is stable against perturbations that preserve translation symmetry.

The theoretical description of Weyl semimetals in terms of the single-particle Hamiltonian (\ref{hweyl}) does not take into account the electron-electron interactions that are always present to some degree in real materials. This is justified to a first approximation: because of the fast vanishing of the density of states $\rho(\varepsilon)\propto\varepsilon^2$ of a Weyl semimetal at the Fermi energy $\varepsilon=0$, short-range interactions are perturbatively irrelevant and Coulomb interactions are marginally irrelevant. The effect of weak interactions can thus be treated in perturbation theory, and leads for example to finite or logarithmic renormalizations of transport properties.\cite{hosur2012,jho2013,rosenstein2013,mastropietro2013} On the other hand, sufficiently strong interactions can lead to a quantum critical point at finite interaction strength where the Weyl semimetal is destroyed. The most likely scenario is that of spontaneous symmetry breaking. Previous theoretical studies have considered specific examples of broken-symmetry states that may occur as a result of strong density-density interactions in a Weyl semimetal, including excitonic and charge-density wave (CDW) ground states,\cite{zyuzin2012,wei2012,wang2013} as well as superconducting ground states.\cite{cho2012,wei2013} These studies begin with a particular microscopic interaction on the lattice that is projected onto the low-energy subspace of Weyl points. The resulting low-energy continuum field theory of interacting Weyl fermions is then studied in the mean-field approximation, assuming a particular decoupling channel. However, constructing a low-energy effective theory requires a somewhat arbitrary choice of high-energy cutoff $\Lambda$, and an effective theory with cutoff $\Lambda$ can in principle be obtained from an effective theory with a different cutoff $\Lambda'>\Lambda$ by integrating out all degrees of freedom with energies between $\Lambda$ and $\Lambda'$. This procedure will generate interaction terms that were absent in the initial projection of the microscopic interaction onto the low-energy subspace. In principle, one should therefore include in the low-energy effective theory all interaction terms that are consistent with the symmetries of the problem.

The first question is whether one should use the symmetry group of the microscopic Hamiltonian on the lattice, or the (larger) symmetry group of the noninteracting Weyl fermion Hamiltonian, e.g., Eq.~(\ref{hweyl}). If the interaction strength is comparable to the bandwidth, such that in perturbation theory the interaction will cause significant mixing between the low-energy Weyl fermions and high-energy states, it is preferable to use the lattice symmetry group. However, if the interaction strength is small compared to the bandwidth, in perturbation theory the low-energy Weyl fermions interact mostly with each other without significant mixing with high-energy states, and it is sensible to constrain the interaction terms by the symmetry group of the noninteracting Weyl fermions. Furthermore, the lattice symmetry group is material-specific whereas the low-energy symmetry group is (almost) universal. Given the diversity of materials that have been predicted to realize the Weyl semimetal, it is useful to focus on those symmetries that are common to the low-energy subspace of many Weyl semimetals, rather than on those that differ from material to material.

In this paper we develop a minimal low-energy description of interacting fermions in a model of Weyl semimetal based on a small set of low-energy symmetries, and investigate its possible broken-symmetry ground states via the renormalization group (RG) method. Starting from a simple model of a Weyl semimetal  on the cubic lattice with two Weyl points,\cite{yang2011} we construct a low-energy effective theory for noninteracting Weyl fermions and determine its symmetry group (Sec.~\ref{sec:noninteracting}). We restrict ourselves to short-range interactions.  Although long-range Coulomb interactions are expected to dominate in potential solid-state realizations of Weyl semimetals where the fermions are charged electrons, besides its academic interest the study of short-range interactions is relevant to other potential realizations of Weyl semimetals where the fermions are electrically neutral, such as ultracold atomic gases in optical lattices with artificial gauge fields\cite{jiang2012,ganeshan2014} or Weyl superconductors.\cite{meng2012} We determine the most general short-range interaction term that is consistent with the symmetry group of the noninteracting low-energy Hamiltonian (Sec.~\ref{sec:SRintSymm}). These symmetry considerations as well as the use of Fierz identities reduce the number of independent coupling constants from 136 to four. We then perform a one-loop RG analysis that determines the flow of coupling constants in this four-dimensional parameter space (Sec.~\ref{sec:RG}). We find a single stable (Gaussian) fixed point, corresponding to the noninteracting Weyl semimetal, as well as four critical points, six bicritical points, and four tricritical points. We are interested in broken-symmetry states that correspond to stable fixed points at infinity. In order to explore possible broken-symmetry ground states at strong coupling, we investigate the stability of trajectories towards strong coupling in the four-dimensional space of coupling constants and find a single stable trajectory.  As in analogous studies of interacting electrons on the honeycomb lattice,\cite{herbut2009} this analysis is admittedly uncontrolled in that the perturbative RG flow is extrapolated to strong coupling, but  has the advantage over previous mean-field studies of not requiring an \emph{a priori} choice of order parameter. We calculate susceptibilities along the stable trajectory and determine the leading symmetry-breaking instability (Sec.~\ref{sec:susceptibilities}), which is a spin-density wave (SDW) with wave vector equal to the momentum-space separation of the Weyl points (Sec.~\ref{sec:LeadingInstability}). The single-particle spectrum in this state is fully gapped. As the Hamiltonian has no spin rotation symmetry, the only Goldstone mode of this generally incommensurate SDW is the sliding mode. The electromagnetic response of the state is unusual in that this sliding mode couples to external electric and magnetic fields like the pseudoscalar axion field of particle physics,\cite{peccei1977} which was also found to occur if the Weyl semimetal develops CDW order.\cite{wang2013} In Sec.~\ref{sec:maximal}, we consider a model with additional symmetries that has a single independent coupling constant. Depending on the sign of the flow to strong coupling, we find gapless ferromagnetic states or fully gapped Fulde-Ferrell-Larkin-Ovchinnikov\cite{fulde1964,larkin1965} (FFLO) superconducting states.

\section{Noninteracting Hamiltonian}
\label{sec:noninteracting}

In this section we derive the low-energy effective field theory for the simplest type of Weyl semimetal with broken time-reversal symmetry, which has two Weyl points related by inversion symmetry. We determine the symmetries of the resulting Lagrangian, which are then used in Sec.~\ref{sec:SRintSymm} to constrain the form of the short-range interaction terms.

\subsection{Lattice model and low-energy effective theory}

The starting point of our analysis is a simple two-band model on the 3D cubic lattice at half-filling,\cite{yang2011}
\begin{align}\label{H0}
H_0=\int\frac{d^3k}{(2\pi)^3}c_{\b{k}\alpha}^\dag h_{\alpha\beta}(\b{k})c_{\b{k}\beta},
\end{align}
where $c_{\b{k}\alpha}^\dag$ ($c_{\b{k}\alpha}$) creates (annihilates) a  fermion with momentum $\b{k}=(k_x,k_y,k_z)$ and spin $\alpha=\uparrow,\downarrow$, the integration is over the first Brillouin zone $(-\pi,\pi)^3$, and the $2\times 2$ Bloch Hamiltonian matrix is
\begin{align}
h(\b{k})&=t(\sigma_1\sin k_x+\sigma_2\sin k_y)+t_z(\cos k_z-\cos Q)\sigma_3\nonumber\\
&\hspace{4mm}+m(2-\cos k_x-\cos k_y)\sigma_3,
\end{align}
where $\sigma_1,\sigma_2,\sigma_3$ are the three Pauli matrices, and $t,t_z,m$ are real parameters. Because time-reversal symmetry (TRS) flips the spin of the { fermion} $\bsigma\rightarrow-\bsigma$ as well as its momentum $\b{k}\rightarrow-\b{k}$, $H_0$ manifestly breaks TRS. This Hamiltonian describes a Weyl semimetal with two Weyl points at $\b{P}_\pm=(0,0,\pm Q)$, where we assume that $Q\neq 0,\pi$. At half-filling, the chemical potential $\mu$ is zero, and the Fermi surface consists of the two Weyl points alone.

Although the physics at energies far from the Weyl points will in general depend on the details of the lattice model one chooses, the physics at energies near the Weyl points only depends on a few parameters. To capture these simple low-energy properties, we derive an effective continuum field theory valid for energies close to the Fermi energy, i.e., near the Weyl points. To do this, we expand the { fermion} operator near the Weyl points,
\begin{align}\label{ExpandElectron}
c_{\b{r}\alpha}&\simeq\left(\int_{|\b{k}-\b{P}_+|<\Lambda}
\frac{d^3k}{(2\pi)^3}+
\int_{|\b{k}-\b{P}_-|<\Lambda}
\frac{d^3k}{(2\pi)^3}\right)e^{i\b{k}\cdot\b{r}}c_{\b{k}\alpha}\nonumber\\
&=e^{i\b{P}_+\cdot\b{r}}\int_{|\b{p}|<\Lambda}\frac{d^3p}{(2\pi)^3}e^{i\b{p}\cdot\b{r}}c_{\b{p}+\b{P}_+,\alpha}\nonumber\\
&\hspace{4mm}+e^{i\b{P}_-\cdot\b{r}}\int_{|\b{p}|<\Lambda}\frac{d^3p}{(2\pi)^3}e^{i\b{p}\cdot\b{r}}c_{\b{p}+\b{P}_-,\alpha}\nonumber\\
&=e^{iQz}\psi_{R\alpha}(\b{r})+e^{-iQz}\psi_{L\alpha}(\b{r}),
\end{align}
where $c_{\b{r}\alpha}=\int\frac{d^3k}{(2\pi)^3}e^{i\b{k}\cdot\b{r}}c_{\b{k}\alpha}$ annihilates a { fermion} on lattice site $\b{r}$, $\Lambda$ is a large-momentum cutoff such that $\Lambda\ll|Q|$, and we define the slow chiral or Weyl fermion fields,
\begin{align}
\psi_{R\alpha}(\b{r})&=\int_{|\b{p}|<\Lambda}\frac{d^3p}{(2\pi)^3}e^{i\b{p}\cdot\b{r}}c_{\b{p}+\b{P}_+,\alpha},\label{psiR}\\
\psi_{L\alpha}(\b{r})&=\int_{|\b{p}|<\Lambda}\frac{d^3p}{(2\pi)^3}e^{i\b{p}\cdot\b{r}}c_{\b{p}+\b{P}_-,\alpha},\label{psiL}
\end{align}
and their Fourier components $\psi_{R\alpha}(\b{p})=c_{\b{p}+\b{P}_+,\alpha}$ and $\psi_{L\alpha}(\b{p})=c_{\b{p}+\b{P}_-,\alpha}$, with $|\b{p}|<\Lambda$. Substituting the expansion (\ref{ExpandElectron}) in the Hamiltonian (\ref{H0}) and expanding $h(\b{k})$ near the Weyl points as well, we obtain
\begin{align}\label{H0cont1}
H_0&\simeq\int_{|\b{p}|<\Lambda}\frac{d^3p}{(2\pi)^3}\Bigl(\psi_{R}^\dag(\b{p})
h_R(\b{p})\psi_{R}(\b{p})\nonumber\\
&\hspace{22mm}+\psi_{L}^\dag(\b{p})
h_L(\b{p})\psi_{L}(\b{p})\Bigr),
\end{align}
where $\psi_R=(\psi_{R\uparrow},\psi_{R\downarrow})$ and $\psi_L=(\psi_{L\uparrow},\psi_{L\downarrow})$ are two-component Weyl spinors, and the $2\times 2$ Weyl Hamiltonians $h_R,h_L$ are
\begin{align}
h_R(\b{p})&=t(\sigma_1 p_x+\sigma_2 p_y)-t_z\sin Q\sigma_3 p_z,\\
h_L(\b{p})&=t(\sigma_1 p_x+\sigma_2 p_y)+t_z\sin Q\sigma_3 p_z,
\end{align}
to leading order in $\b{p}$. The chirality $c$, given by $c=\sgn(\b{v}_1\cdot(\b{v}_2\times\b{v}_3))$ for a Hamiltonian of the form $h\sim\sum_i \b{v}_i\cdot\b{p}\sigma_i$,\cite{wan2011} is $c_R=\sgn(-t^2t_z\sin Q)=-c_L$ and thus opposite for each Weyl point. Defining $v_\parallel=t$ and $v_z=-t_z\sin Q$, Eq.~(\ref{H0cont1}) can be written as
\begin{align}\label{H0Dirac}
H_0=\int\frac{d^3p}{(2\pi)^3}\Psi^\dag(\b{p})
\left(v_\parallel\b{p}_\parallel\cdot
\boldsymbol{\Gamma}_\parallel+v_zp_z\Gamma_3\right)
\Psi(\b{p}),
\end{align}
where $|\b{p}|<\Lambda$ is assumed. We define the four-component Dirac spinor,
\begin{align}\label{DiracSpinor}
\Psi=\left(\begin{array}{c}
\psi_R\\
\psi_L
\end{array}\right),
\end{align}
with $\b{p}_\parallel=(p_x,p_y)$ and $\boldsymbol{\Gamma}_\parallel=(\Gamma_1,\Gamma_2)$. We define the five Hermitian gamma matrices
\begin{align}
&\Gamma_1=\tau_0\otimes\sigma_1,\hspace{5mm}
\Gamma_2=\tau_0\otimes\sigma_2,\hspace{5mm}
\Gamma_3=\tau_3\otimes\sigma_3,\nonumber\\
&\Gamma_4=\tau_1\otimes\sigma_3,\hspace{5mm}
\Gamma_5=\tau_2\otimes\sigma_3,
\end{align}
where $\tau_1,\tau_2,\tau_3$ are Pauli matrices acting in the space of Weyl points, and $\tau_0$ is the $2\times 2$ identity matrix. These matrices obey the $SO(5)$ Clifford algebra $\{\Gamma_a,\Gamma_b\}=2\delta_{ab}$, $a,b=1,\ldots,5$. For future use we  define the ten additional Hermitian matrices
\begin{align}
\Gamma_{ab}=\frac{1}{2i}[\Gamma_a,\Gamma_b],\hspace{5mm}
a,b=1,\ldots,5,\hspace{5mm}a<b,
\end{align}
that also square to the identity, and form a complete set of generators of the $so(5)$ Lie algebra.\cite{murakami2004} The set of fifteen traceless Hermitian matrices $\Gamma_a,\Gamma_{ab}$ generates the $su(4)$ Lie algebra, and denoting the $4\times 4$ identity matrix by $\Gamma_0$, the set of sixteen Hermitian matrices $\Gamma_0,\Gamma_a,\Gamma_{ab}$ is a complete basis for all $4\times 4$ Hermitian matrices. This latter fact will be useful in our construction of short-range interaction terms in Sec.~\ref{sec:SRintSymm}.

The low-energy effective Hamiltonian (\ref{H0Dirac}) is the massless Dirac Hamiltonian in 3+1 dimensions, and the energy spectrum is gapless with linearly dispersing positive ($+$) and negative ($-$) energy branches $E_\pm(\b{p})=\pm\sqrt{v_\parallel^2\b{p}_\parallel^2
+v_z^2p_z^2}$ meeting at $\b{p}=0$. In the following subsections, we determine the symmetries of this effective Hamiltonian. We denote symmetry operators acting in the many-body Hilbert space by curly letters $\mathcal{S}$ and finite-dimensional, unitary representation matrices by regular letters $S$.

\subsection{Discrete symmetries}
\label{sec:DiscreteSymm}

\subsubsection{Parity symmetry}

There must exist an inversion or parity symmetry that interchanges the two Weyl points. We therefore define a unitary parity operator $\mathcal{P}$ that obeys $\mathcal{P}^2=1$ by its action on the Weyl spinors (\ref{psiR})-(\ref{psiL}),
\begin{align}
\mathcal{P}\psi_R(\b{r})\mathcal{P}^{-1}&=\tilde{P}\psi_L(-\b{r}),\label{InversionR}\\
\mathcal{P}\psi_L(\b{r})\mathcal{P}^{-1}&=\tilde{P}\psi_R(-\b{r}),\label{InversionL}
\end{align}
where the $2\times 2$ representation matrix $\tilde{P}$ must satisfy $\tilde{P}^2=1$. One can directly check that $H_0$ commutes with $\mathcal{P}$ if $h_R(\b{p})=\tilde{P}^Th_L(-\b{p})\tilde{P}$ and $h_L(\b{p})=\tilde{P}^Th_R(-\b{p})\tilde{P}$, which is satisfied by the choice $\tilde{P}=\sigma_3$. Parity therefore acts on the Dirac spinor (\ref{DiracSpinor}) as
\begin{align}
\mathcal{P}\Psi(\b{r})\mathcal{P}^{-1}=P\Psi(-\b{r}),
\end{align}
where the $4\times 4$ representation matrix $P$, which also obeys $P^2=1$, is
\begin{align}
P=\tau_1\otimes\sigma_3=\Gamma_4.
\end{align}

\subsubsection{Antiunitary symmetry}

The original lattice model (\ref{H0}) breaks the physical TRS. However, the effective Hamiltonian (\ref{H0Dirac}) commutes with the antiunitary operator $\mathcal{T}$ that satisfies $\mathcal{T}^2=-1$ and is defined by
\begin{align}\label{EqforT}
\mathcal{T}\Psi(\b{r})\mathcal{T}^{-1}=TK\Psi(\b{r}),
\end{align}
where $K$ denotes complex conjugation of $c$-numbers and the $4\times 4$ representation matrix $T$, which obeys $T^2=-1$, is
\begin{align}
T=\tau_0\otimes i\sigma_2=i\Gamma_2.
\end{align}
This antiunitary symmetry does not interchange the two Weyl points, but flips the spin of a { fermion} near a given Weyl point.

\subsubsection{Particle-hole symmetry}

It is known in the context of relativistic quantum field theory that the free massless Dirac fermion in 3+1 dimensions is invariant under charge conjugation. Likewise, assuming that the Hamiltonian (\ref{H0Dirac}) is normal ordered $H_0\equiv\colon H_0\colon$ with respect to the creation and annihilation operators $\Psi^\dag,\Psi$, it commutes with a unitary particle-hole symmetry operator $\mathcal{C}$ that obeys $\mathcal{C}^2=1$ and is defined by
\begin{align}\label{EqforC}
\mathcal{C}\Psi(\b{r})\mathcal{C}^{-1}=C(\Psi^\dag(\b{r}))^T,
\end{align}
where the $4\times 4$ representation matrix $C$, which obeys $C^2=1$, is
\begin{align}
C=\tau_0\otimes\sigma_2=\Gamma_2.
\end{align}
The particle-hole symmetry does not interchange the two Weyl points.

\subsection{Continuous symmetries}
\label{sec:ContSymm}

The Hamilonian (\ref{H0Dirac}) is manifestly invariant under translation symmetry $\mathcal{T}(\b{R})\Psi(\b{r})\mathcal{T}(\b{R})^{-1}=\Psi(\b{r}+\b{R})$, where $\mathcal{T}(\b{R})$ is the unitary operator for translation in real space by the vector $\b{R}$, and under the $U(1)$ global symmetry $\mathcal{G}(\alpha)\Psi(\b{r})\mathcal{G}(\alpha)^{-1}=e^{-i\alpha}\Psi(\b{r})$, where $\mathcal{G}(\alpha)$ is the unitary operator for position and time-independent gauge transformations by a phase $\alpha$. The interaction terms we will consider are also manifestly invariant under these symmetries, hence the latter will not constrain the form of the former.

\subsubsection{Rotation symmetry}
\label{sec:RotSymm}

Because of the anisotropy in the velocities $v_\parallel\neq v_z$, the Hamiltonian (\ref{H0Dirac}) does not have a full $SO(3)$ rotation symmetry but rather an $SO(2)$ symmetry under rotations about the axis joining the two Weyl points (here the $z$ axis). Because the spin and orbital angular momenta are mixed by the Hamiltonian, this $SO(2)$ symmetry corresponds to the conservation of total angular momentum in the $z$ direction, $J_z=L_z+\frac{1}{2}\sigma_3$, where $L_z=-i\frac{\partial}{\partial\varphi}$ is the orbital angular momentum with $\varphi$ the azimuthal angle in the $xy$ plane. We have $[\mathcal{R}(\theta),H_0]=0$ where the unitary rotation operator $\mathcal{R}(\theta)$ is defined by
\begin{align}
\mathcal{R}(\theta)\Psi(\b{r})\mathcal{R}(\theta)^{-1}=R(\theta)\Psi(R_\theta\b{r}),\hspace{5mm}
0\leq\theta<2\pi.
\end{align}
The $4\times 4$ representation matrix $R(\theta)$ is
\begin{align}\label{Rtheta}
R(\theta)=e^{-i\theta\Gamma_{12}/2},
\end{align}
and $R_\theta$ is an $SO(2)$ rotation matrix that acts only on the $x$ and $y$ coordinates,
\begin{align}\label{SO2RotationMatrix}
R_\theta=\left(\begin{array}{ccc}
\cos\theta & -\sin\theta & 0 \\
\sin\theta & \cos\theta & 0 \\
0 & 0 & 1
\end{array}\right).
\end{align}
The rotation symmetry does not interchange the two Weyl points.

\subsubsection{Chiral symmetry}
\label{sec:ChiralSymm}

The free massless Dirac fermion in 3+1 dimensions is invariant under a $U(1)$ chiral symmetry that describes the fact that in the absence of electromagnetic fields, the currents of right-handed and left-handed Weyl fermions are separately conserved. Mathematically, we have $[\mathcal{R}_\chi(\phi),H_0]=0$ where the unitary chiral symmetry operator $\mathcal{R}_\chi(\phi)$ is defined by
\begin{align}
\mathcal{R}_\chi(\phi)\Psi(\b{r})\mathcal{R}_\chi(\phi)^{-1}
=R_\chi(\phi)\Psi(\b{r}),\hspace{5mm}\phi\leq 0<2\pi,
\end{align}
where the $4\times 4$ representation matrix $R_\chi(\phi)$ is
\begin{align}\label{Rchiral}
R_\chi(\phi)=e^{-i\phi\Gamma_{45}/2}.
\end{align}

\subsection{Additional chiral symmetries}
\label{sec: additional}

The noninteracting Hamiltonian $H_0$ has additional chiral symmetries besides the $U(1)$ chiral ``charge'' symmetry of Sec.~\ref{sec:ChiralSymm}. Indeed, because $H_0$ only contains block-diagonal gamma matrices, the right-handed and left-handed Weyl fermions are completely decoupled, and we can in principle define separate antiunitary symmetries $\mathcal{T}_R,\mathcal{T}_L$, particle-hole symmetries $\mathcal{C}_R,\mathcal{C}_L$, and rotation symmetries $\mathcal{R}_R(\theta),\mathcal{R}_L(\theta)$ for each of those. Alternatively, we can define chiral versions $\tilde{\mathcal{T}},\tilde{\mathcal{C}},\tilde{\mathcal{R}}(\theta)$ of the symmetries $\mathcal{T},\mathcal{C},\mathcal{R}(\theta)$ we have already discussed that are additional symmetries of $H_0$. A chiral antiunitary symmetry $\tilde{\mathcal{T}}$ with $\tilde{\mathcal{T}}^2=-1$ can be defined as in Eq.~(\ref{EqforT}) but where $\tilde{T}=\tau_3\otimes i\sigma_2=-i\Gamma_{13}$. Likewise, a chiral particle-hole symmetry $\tilde{\mathcal{C}}$ with $\tilde{\mathcal{C}}^2=1$ can be defined as in Eq.~(\ref{EqforC}) but with $\tilde{C}=\tau_3\otimes\sigma_2=-\Gamma_{13}$. Independent rotations of two Weyl points, which form the group $SO(2)_R\times SO(2)_L$, can be divided into normal $SO(2)_{R+L}$ rotations generated by $\tau_0\otimes\sigma_3=\Gamma_{12}$ (Sec.~\ref{sec:RotSymm}), and chiral $SO(2)_{R-L}$ rotations generated by $\tau_3\otimes\sigma_3=\Gamma_3$. However, these additional chiral symmetries will not be respected by most lattice-scale interactions, as we now discuss. 

\section{Short-range interactions and symmetries}
\label{sec:SRintSymm}

The simplest type of interactions one can consider adding to the noninteracting Hamiltonian discussed in the previous section are short-range interactions. In the lattice model (\ref{H0}), the first choice that comes to mind is the on-site Hubbard interaction,
\begin{align}\label{hubbard}
V=U\sum_\b{r}n_{\b{r}\uparrow}n_{\b{r}\downarrow}
=\frac{U}{2}\sum_{\b{r}}c_{\b{r}\alpha}^\dag c_{\b{r}\alpha}c_{\b{r}\beta}^\dag c_{\b{r}\beta},
\end{align}
where we have ignored a one-body term that can be absorbed in a redefinition of the chemical potential. In the low-energy limit, we can substitute the expression (\ref{ExpandElectron}) for the { fermion} operator into Eq.~(\ref{hubbard}). Because $\psi_R$ and $\psi_L$ are slow fields with Fourier components much less than $|Q|$, terms containing $e^{\pm 2iQz}$ will average out to zero in the integral over $\b{r}$. We obtain
\begin{align}\label{HubbardLowEnergy}
V\simeq\frac{U}{2}\int d^3r\left(\rho_R^2+\rho_L^2+2\rho_R\rho_L
+2\psi_{R\alpha}^\dag\psi_{L\alpha}
\psi_{L\beta}^\dag\psi_{R\beta}\right),
\end{align}
where we define the chiral density operators $\rho_R=\psi_{R\alpha}^\dag\psi_{R\alpha}$ and $\rho_L=\psi_{L\alpha}^\dag\psi_{L\alpha}$. This effective Hubbard interaction respects all the symmetries of the noninteracting Hamiltonian, except the additional chiral symmetries enumerated in Sec.~\ref{sec: additional}. We expect this to be a generic feature of interactions: because lattice-scale interactions are capable of scattering particles between Weyl points, we do not expect to be able to define separate right and left symmetries once interactions are incorporated. We thus ignore the additional chiral symmetries, operating on the assumption that these are broken by interactions. We discuss in Sec.~\ref{sec:maximal} the toy model that results if the additional chiral symmetries are respected by the interactions. 

While the Hubbard interaction is a natural first guess, we want to study the most general possible interaction Hamiltonian, subject to some symmetry constraints that we will shortly discuss. To this end, we note that each term in Eq.~(\ref{HubbardLowEnergy}) is of the form $\int d^3r[\Psi^\dag(\b{r})M_1\Psi(\b{r})][\Psi^\dag(\b{r})M_2\Psi(\b{r})]$ where $M_1$ and $M_2$ are constant $4\times 4$ Hermitian matrices. We are interested in the most general short-range interaction term, which will contain all possible such terms (the requirement that $M_1$ and $M_2$ be Hermitian comes solely from the requirement that $V$ be a Hermitian operator). As mentioned before, any Hermitian $4\times 4$ matrix can be expanded in the basis of the sixteen Hermitian matrices $\Gamma_A\in\{\Gamma_0,\Gamma_a,\Gamma_{ab}\}$. Therefore the most general short-range interaction term is
\begin{align}
V=\int d^3r\,g_{AB}(\Psi^\dag\Gamma_A\Psi)(\Psi^\dag\Gamma_B\Psi),
\end{align}
where $g_{AB}$ is a real symmetric $16\times 16$ matrix that has $(16\times 17)/2=136$ independent entries. However, the number of independent interactions can be drastically reduced by demanding that $V$ be invariant under the symmetries of the noninteracting Hamiltonian $H_0$, discussed in Sec.~\ref{sec:DiscreteSymm} and Sec.~\ref{sec:ContSymm}. To implement this program we follow an approach used previously to study interacting electrons in graphene.\cite{herbut2009,vafek2010b}

\subsection{Parity symmetry}

Under parity $\mathcal{P}$, a typical interaction term $\int d^3r(\Psi^\dag\Gamma_A\Psi)(\Psi^\dag\Gamma_B\Psi)$ transforms as
\begin{align}
&\mathcal{P}\int d^3r(\Psi^\dag\Gamma_A\Psi)(\Psi^\dag\Gamma_B\Psi)\mathcal{P}^{-1}\nonumber\\
&=\int d^3r\left(\Psi^\dag(-\b{r})P^{-1}\Gamma_AP\Psi(-\b{r})\right)
\nonumber\\
&\hspace{10mm}\times
\left(\Psi^\dag(-\b{r})P^{-1}\Gamma_BP\Psi(-\b{r})\right)\nonumber\\
&=\int d^3r\left(\Psi^\dag(\b{r})P^{-1}\Gamma_AP\Psi(\b{r})\right)
\left(\Psi^\dag(\b{r})P^{-1}\Gamma_BP\Psi(\b{r})\right),
\end{align}
hence invariance under parity requires that either both $\Gamma_A$ and $\Gamma_B$ are even under parity ($[P,\Gamma_A]=[P,\Gamma_B]=0$), or both $\Gamma_A$ and $\Gamma_B$ are odd under parity ($\{P,\Gamma_A\}=\{P,\Gamma_B\}=0$). Using the gamma matrix identities listed in Appendix~\ref{app:gamma}, we find that the eight matrices $A_i\in\{\Gamma_0,\Gamma_4,\Gamma_{12},\Gamma_{13},\Gamma_{15},\Gamma_{23},\Gamma_{25},\Gamma_{35}\}$, $i=1,\ldots,8$ are even under parity, while the remaining eight matrices $B_j\in\{\Gamma_1,\Gamma_2,\Gamma_3,\Gamma_5,\Gamma_{14},\Gamma_{24},\Gamma_{34},\Gamma_{45}\}$, $j=1,\ldots,8$ are odd. The interaction Hamiltonian therefore becomes
\begin{align}\label{Vparity}
V=\int d^3r\left(a_{ij}(\Psi^\dag A_i\Psi)(\Psi^\dag A_j\Psi)+b_{ij}(\Psi^\dag B_i\Psi)(\Psi^\dag B_j\Psi)\right),
\end{align}
where $a_{ij}$ and $b_{ij}$ are real symmetric $8\times 8$ matrices with $(8\times 9)/2=36$ independent couplings each, for a total of $2\times 36=72$ independent couplings.

\subsection{Rotation symmetry}
\label{sec:SRintRot}

Under rotation $\mathcal{R}(\theta)$, an interaction term transforms as
\begin{align}
&\mathcal{R}(\theta)\int d^3r(\Psi^\dag\Gamma_A\Psi)(\Psi^\dag\Gamma_B\Psi)\mathcal{R}(\theta)^{-1}\nonumber\\
&=\int d^3r\left(\Psi^\dag(\b{r}')R(\theta)^{-1}\Gamma_AR(\theta)
\Psi(\b{r}')\right)\nonumber\\
&\hspace{10mm}\times\left(\Psi^\dag(\b{r}')R(\theta)^{-1}\Gamma_BR(\theta)
\Psi(\b{r}')\right)\nonumber\\
&=\int d^3r'\left(\Psi^\dag(\b{r}')R(\theta)^{-1}\Gamma_AR(\theta)
\Psi(\b{r}')\right)\nonumber\\
&\hspace{10mm}\times\left(\Psi^\dag(\b{r}')R(\theta)^{-1}\Gamma_BR(\theta)
\Psi(\b{r}')\right),
\end{align}
where $\b{r}'=R_\theta\b{r}$. Invariance under rotation requires that either both $\Gamma_A$ and $\Gamma_B$ are scalars under rotations: $[\Gamma_{12},\Gamma_A]=[\Gamma_{12},\Gamma_B]=0$ [see Eq.~(\ref{Rtheta})], or the interaction term has to be of the ``dot-product'' form $(\Psi^\dag\eta_i\Psi)(\Psi^\dag\eta'_i\Psi)$ or ``cross-product'' form $\epsilon_{ij}(\Psi^\dag\eta_i\Psi)(\Psi^\dag\eta'_j\Psi)$ where $\boldsymbol{\eta}=(\eta_1,\eta_2)$ is a pair of gamma matrices that transform as a vector under rotations: $[\Gamma_{12},\eta_i]=\pm 2i\epsilon_{ij}\eta_j$ and $[\Gamma_{12},\eta_i']=\pm 2i\epsilon_{ij}\eta_j'$. We find that the eight gamma matrices $\Gamma_0,\Gamma_3,\Gamma_4,\Gamma_5,\Gamma_{12},\Gamma_{34},\Gamma_{35},\Gamma_{45}$ are scalars, while the remaining eight form four vectors:
\begin{align}
&\boldsymbol{\alpha}=(\Gamma_1,\Gamma_2),\hspace{5mm} \boldsymbol{\beta}=(\Gamma_{13},\Gamma_{23}),\nonumber\\ &\boldsymbol{\gamma}=(\Gamma_{14},\Gamma_{24}),\hspace{5mm}
\boldsymbol{\delta}=(\Gamma_{15},\Gamma_{25}),
\end{align}
such that $[\Gamma_{12},\alpha_i]=2i\epsilon_{ij}\alpha_j$, and similarly for $\boldsymbol{\beta},\boldsymbol{\gamma},\boldsymbol{\delta}$.

However, we have to respect the structure (\ref{Vparity}) already imposed by parity symmetry. The four scalars $\Gamma_0,\Gamma_4,\Gamma_{12},\Gamma_{35}$ are even under parity while the other four $\Gamma_3,\Gamma_5,\Gamma_{34},\Gamma_{45}$ are odd. For the vectors, we find that $\boldsymbol{\beta}$ and $\boldsymbol{\delta}$ are even under parity while $\boldsymbol{\alpha}$ and $\boldsymbol{\gamma}$ are odd. We denote the even-parity scalars by $\Gamma_i^{(e)}=\{\Gamma_0,\Gamma_4,\Gamma_{12},\Gamma_{35}\}$, $i=1,\ldots,4$, the odd-parity scalars by $\Gamma_i^{(o)}=\{\Gamma_3,\Gamma_5,\Gamma_{34},\Gamma_{45}\}$, $i=1,\ldots,4$, the even-parity vectors by $\boldsymbol{\Gamma}^{(e)}_i=\{\boldsymbol{\beta},\boldsymbol{\delta}\}$, $i=1,2$, and the odd-parity vectors by $\boldsymbol{\Gamma}^{(o)}_i=\{\boldsymbol{\alpha},\boldsymbol{\gamma}\}$, $i=1,2$. (In the previous sentence, $\{,\}$ does not denote the anticommutator but simply a set.) With this notation, the interaction term becomes
\begin{align}\label{Vrot}
V&=\int d^3r\nonumber\\
&\times\Bigl[g_{ij}^{(e)}(\Psi^\dag\Gamma_i^{(e)}\Psi)(\Psi^\dag\Gamma_j^{(e)}\Psi)+g_{ij}^{(o)}(\Psi^\dag\Gamma_i^{(o)}\Psi)(\Psi^\dag\Gamma_j^{(o)}\Psi)\nonumber\\
&+g_{\beta\beta}(\Psi^\dag\boldsymbol{\beta}\Psi)^2
+g_{\delta\delta}(\Psi^\dag\boldsymbol{\delta}\Psi)^2
+g_{\beta\cdot\delta}(\Psi^\dag\boldsymbol{\beta}\Psi)\cdot
(\Psi^\dag\boldsymbol{\delta}\Psi)\nonumber\\
&+g_{\alpha\alpha}(\Psi^\dag\boldsymbol{\alpha}\Psi)^2
+g_{\gamma\gamma}(\Psi^\dag\boldsymbol{\gamma}\Psi)^2
+g_{\alpha\cdot\gamma}(\Psi^\dag\boldsymbol{\alpha}\Psi)\cdot
(\Psi^\dag\boldsymbol{\gamma}\Psi)\nonumber\\
&+g_{\beta\times\delta}(\Psi^\dag\boldsymbol{\beta}\Psi)\times
(\Psi^\dag\boldsymbol{\delta}\Psi)
+g_{\alpha\times\gamma}(\Psi^\dag\boldsymbol{\alpha}\Psi)\times
(\Psi^\dag\boldsymbol{\gamma}\Psi)
\Bigr],
\end{align}
where $g_{ij}^{(e)}$ and $g_{ij}^{(o)}$ are real symmetric $4\times 4$ matrices with $(4\times 5)/2=10$ independent couplings each, so that we have a total of $2\times 10+8=28$ independent couplings.

\subsection{Antiunitary symmetry}

Under the antiunitary symmetry $\mathcal{T}$, an interaction term transforms as
\begin{align}
&\mathcal{T}\int d^3r(\Psi^\dag\Gamma_A\Psi)(\Psi^\dag\Gamma_B\Psi)\mathcal{T}^{-1}\nonumber\\
&=\int d^3r(\Psi^\dag T^{-1}\Gamma_A^*T\Psi)
(\Psi^\dag T^{-1}\Gamma_B^*T\Psi),
\end{align}
hence invariance under the antiunitary symmetry requires that either both $\Gamma_A$ and $\Gamma_B$ are even ($T\Gamma_AT^{-1}=\Gamma_A^*$ and $T\Gamma_BT^{-1}=\Gamma_B^*$), or both $\Gamma_A$ and $\Gamma_B$ are odd ($T\Gamma_AT^{-1}=-\Gamma_A^*$ and $T\Gamma_BT^{-1}=-\Gamma_B^*$) under this symmetry. We find that the six matrices $\Gamma_0,\Gamma_5,\Gamma_{15},\Gamma_{25},\Gamma_{35},\Gamma_{45}$ are even, while the remaining ten matrices $\Gamma_1,\Gamma_2,\Gamma_3,\Gamma_4,\Gamma_{12},\Gamma_{13},\Gamma_{14},\Gamma_{23},\Gamma_{24},\Gamma_{34}$ are odd. We subdivide the matrices $\Gamma_i^{(e)},\Gamma_i^{(o)},\boldsymbol{\Gamma}^{(e)}_i,\boldsymbol{\Gamma}^{(o)}_i$ of Sec.~\ref{sec:SRintRot} into matrices that are even ($+$) or odd ($-$) under $T$:
\begin{align}
\Gamma_i^{(e,+)}&=\{\Gamma_0,\Gamma_{35}\},\hspace{3mm}\Gamma_i^{(e,-)}=\{\Gamma_4,\Gamma_{12}\},\hspace{2mm}i=1,2,\nonumber\\
\Gamma_i^{(o,+)}&=\{\Gamma_5,\Gamma_{45}\},\hspace{3mm}\Gamma_i^{(o,-)}=\{\Gamma_3,\Gamma_{34}\},\hspace{2mm}i=1,2,\nonumber\\
\boldsymbol{\Gamma}^{(e,+)}&=\boldsymbol{\delta}=(\Gamma_{15},\Gamma_{25}),\hspace{3mm}
\boldsymbol{\Gamma}^{(e,-)}=\boldsymbol{\beta}
=(\Gamma_{13},\Gamma_{23}),\nonumber\\
\boldsymbol{\Gamma}_i^{(o,-)}&=\{\boldsymbol{\alpha},\boldsymbol{\gamma}\}=\{(\Gamma_1,\Gamma_2),(\Gamma_{14},\Gamma_{24})\},\hspace{2mm}i=1,2,
\end{align}
where $\{,\}$ does not denote the anticommutator but simply a set. Respecting the structure of Eq.~(\ref{Vrot}), the interaction term becomes
\begin{align}\label{Vantiunitary}
V&=\int d^3r\Bigl[g_{ij}^{(e,+)}(\Psi^\dag\Gamma_i^{(e,+)}\Psi)(\Psi^\dag\Gamma_j^{(e,+)}\Psi)\nonumber\\
&+g_{ij}^{(e,-)}(\Psi^\dag\Gamma_i^{(e,-)}\Psi)(\Psi^\dag\Gamma_j^{(e,-)}\Psi)\nonumber\\
&+g_{ij}^{(o,+)}(\Psi^\dag\Gamma_i^{(o,+)}\Psi)(\Psi^\dag\Gamma_j^{(o,+)}\Psi)\nonumber\\
&+g_{ij}^{(o,-)}(\Psi^\dag\Gamma_i^{(o,-)}\Psi)(\Psi^\dag\Gamma_j^{(o,-)}\Psi)\nonumber\\
&+g_\delta(\Psi^\dag\boldsymbol{\Gamma}^{(e,+)}\Psi)^2
+g_\beta(\Psi^\dag\boldsymbol{\Gamma}^{(e,-)}\Psi)^2
\nonumber\\
&+g_{\alpha\alpha}(\Psi^\dag\boldsymbol{\alpha}\Psi)^2
+g_{\gamma\gamma}(\Psi^\dag\boldsymbol{\gamma}\Psi)^2\nonumber\\
&+g_{\alpha\cdot\gamma}(\Psi^\dag\boldsymbol{\alpha}\Psi)\cdot
(\Psi^\dag\boldsymbol{\gamma}\Psi)
+g_{\alpha\times\gamma}(\Psi^\dag\boldsymbol{\alpha}\Psi)\times
(\Psi^\dag\boldsymbol{\gamma}\Psi)\Bigr],
\end{align}
where $g_{ij}^{(e,\pm)}$ and $g_{ij}^{(o,\pm)}$ are real symmetric $2\times 2$ matrices with $(2\times 3)/2=3$ independent couplings each, hence we have a total of $4\times 3+6=18$ independent couplings.

\subsection{Particle-hole symmetry}

Under particle-hole symmetry $\mathcal{C}$, assuming that the bilinears appearing in the interaction term are normal ordered, this interaction term transforms as
\begin{align}
&\mathcal{C}\int d^3r\colon\Psi^\dag\Gamma_A\Psi\colon
\colon\Psi^\dag\Gamma_B\Psi\colon\mathcal{C}^{-1}\nonumber\\
&=\int d^3r\colon\Psi^\dag C^{-1}\Gamma_A^TC\Psi\colon
\colon\Psi^\dag C^{-1}\Gamma_B^TC\Psi\colon,
\end{align}
hence invariance under particle-hole symmetry requires that either both $\Gamma_A$ and $\Gamma_B$ are even ($C\Gamma_AC^{-1}=\Gamma_A^T$ and $C\Gamma_BC^{-1}=\Gamma_B^T$), or both $\Gamma_A$ and $\Gamma_B$ are odd ($C\Gamma_AC^{-1}=-\Gamma_A^T$ and $C\Gamma_BC^{-1}=-\Gamma_B^T$) under this symmetry. We find that all the gamma matrices that are even under $T$ are also even under $C$, and all those that are odd under $T$ are also odd under $C$. Therefore particle-hole symmetry does not further reduce the number of independent couplings.

\subsection{Chiral symmetry}
\label{sec:IntChiralSymm}

Under chiral symmetry $\mathcal{R}_\chi(\phi)$, an interaction term transforms as
\begin{align}
&\mathcal{R}_\chi(\phi)\int d^3r(\Psi^\dag\Gamma_A\Psi)(\Psi^\dag\Gamma_B\Psi)\mathcal{R}_\chi(\phi)^{-1}\nonumber\\
&=\int d^3r\left(\Psi^\dag R_\chi(\phi)^{-1}\Gamma_AR_\chi(\phi)
\Psi\right)\nonumber\\
&\hspace{10mm}\times\left(\Psi^\dag R_\chi(\phi)^{-1}\Gamma_BR_\chi(\phi)
\Psi\right).
\end{align}
The analysis is similar to rotation symmetry in Sec.~\ref{sec:SRintRot}. The sixteen gamma matrices divide into chiral scalars that commute with $\Gamma_{45}$ [see Eq.~(\ref{Rchiral})], and chiral vectors $\boldsymbol{\rho}=(\rho_1,\rho_2)$ that satisfy $[\Gamma_{45},\rho_i]=\pm 2i\epsilon_{ij}\rho_j$. We find that the eight gamma matrices $\Gamma_0,\Gamma_1,\Gamma_2,\Gamma_3,\Gamma_{12},\Gamma_{13},\Gamma_{23},\Gamma_{45}$ are chiral scalars, and the remaining eight form four chiral vectors: $(\Gamma_4,\Gamma_5),(\Gamma_{14},\Gamma_{15}),(\Gamma_{24},\Gamma_{25}),(\Gamma_{34},\Gamma_{35})$. Respecting the structure of Eq.~(\ref{Vantiunitary}), the interaction term becomes
\begin{align}\label{VbeforeFierz}
V=\int d^3r\sum_{i=1}^9\lambda_iX_i,
\end{align}
where there are nine independent couplings $\lambda_1,\ldots,\lambda_9$, and the nine quartic terms $X_1,\ldots,X_9$ are
\begin{align}\label{X1X9}
X_1&=(\Psi^\dag\Gamma_0\Psi)^2,\nonumber\\
X_2&=(\Psi^\dag\Gamma_{34}\Psi)^2+(\Psi^\dag\Gamma_{35}\Psi)^2,\nonumber\\
X_3&=(\Psi^\dag\Gamma_4\Psi)^2+(\Psi^\dag\Gamma_5\Psi)^2,\nonumber\\
X_4&=(\Psi^\dag\Gamma_{12}\Psi)^2,\nonumber\\
X_5&=(\Psi^\dag\Gamma_{45}\Psi)^2,\nonumber\\
X_6&=(\Psi^\dag\Gamma_3\Psi)^2,\nonumber\\
X_7&=(\Psi^\dag\Gamma_{14}\Psi)^2
+(\Psi^\dag\Gamma_{24}\Psi)^2+(\Psi^\dag\Gamma_{15}\Psi)^2
+(\Psi^\dag\Gamma_{25}\Psi)^2,\nonumber\\
X_8&=(\Psi^\dag\Gamma_{13}\Psi)^2+(\Psi^\dag\Gamma_{23}\Psi)^2,\nonumber\\
X_9&=(\Psi^\dag\Gamma_1\Psi)^2+(\Psi^\dag\Gamma_2\Psi)^2.
\end{align}

\subsection{Fierz identities}

Although symmetries have reduced the number of independent couplings to nine, not all nine couplings are actually independent because the quartic terms (\ref{X1X9}) are not all linearly independent. Linear relations between products of fermion bilinears are known as Fierz identities. The Fierz identity relevant for our purposes is\cite{herbut2009}
\begin{align}\label{fierzid}
&(\Psi^\dag M\Psi)(\Psi^\dag N\Psi)\nonumber\\
&\hspace{10mm}=-\frac{1}{16}(\Tr M\Gamma_A N\Gamma_B)
(\Psi^\dag\Gamma_B\Psi)(\Psi^\dag\Gamma_A\Psi),
\end{align}
where $M,N$ are arbitrary $4\times 4$ Hermitian matrices, and the sum over $A$ and $B$ is over all sixteen gamma matrices. This identity is proved in Appendix A of Ref.~\onlinecite{herbut2009}.

Using the Fierz identity, we can express the linear dependence of the nine quartic terms in Eq.~(\ref{X1X9}) by the equation $F\b{X}=0$ where we define the column vector $\b{X}=(X_1,\ldots,X_9)$ and the $9\times 9$ Fierz matrix $F$ by
\begin{align}\label{fierzF}
F=\left(\begin{array}{ccccccccc}
5 & 1 & 1 & 1 & 1 & 1 & 1 & 1 & 1 \\
1 & 2 & 0 & 1 & -1 & -1 & 0 & -1 & 1 \\
1 & 0 & 2 & 1 & -1 & -1 & 0 & 1 & -1 \\
1 & 1 & 1 & 5 & 1 & 1 & -1 & -1 & -1 \\
1 & -1 & -1 & 1 & 5 & 1 & -1 & 1 & 1 \\
1 & -1 & -1 & 1 & 1 & 5 & 1 & -1 & -1 \\
1 & 0 & 0 & -1 & -1 & 1 & 1 & 0 & 0 \\
1 & -1 & 1 & -1 & 1 & -1 & 0 & 2 & 0 \\
1 & 1 & -1 & -1 & 1 & -1 & 0 & 0 & 2
\end{array}\right).
\end{align}
By performing Gaussian elimination on $F$, we find that there are only four linearly independent quartic terms amongst the nine. One possible choice of linearly independent quartic terms is $X_1,X_5,X_6,X_9$, and the interaction term reduces to
\begin{align}
V=\int d^3r\left(\lambda_1X_1+\lambda_5X_5
+\lambda_6X_6+\lambda_9X_9\right), \label{eq: realistic}
\end{align}
where the new $\lambda_i$ are linear combinations of the old. For example, the effective Hubbard interaction (\ref{HubbardLowEnergy}) falls into this category, with $(\lambda_1,\lambda_5,\lambda_6,\lambda_9)=(\frac{U}{4},\frac{U}{4},0,-\frac{U}{4})$. 

\section{Renormalization group analysis}
\label{sec:RG}

Our goal is to explore the possible ground states of the Hamiltonian $H=H_0+V$ given by
\begin{align}\label{fullH}
H=\int d^3r&\bigl[\Psi^\dag(
-iv_\parallel\boldsymbol{\Gamma}_\parallel\cdot
\boldsymbol{\partial}_\parallel
-iv_z\Gamma_3\partial_z)\Psi\nonumber\\
&+\lambda_1X_1+\lambda_5X_5
+\lambda_6X_6+\lambda_9X_9\bigr],
\end{align}
using renormalization group (RG) methods. Models of massless Dirac fermions in 3+1 dimensions interacting via short-range four-fermion interactions were first studied in the context of elementary particle physics by Nambu and Jona-Lasinio.\cite{nambu1961} In the absence of interactions, $H_0$ corresponds to the Gaussian fixed point with dynamic critical exponent $z=1$. The short-range interaction term $V$ is perturbatively irrelevant at the Gaussian fixed point, meaning that this term can be neglected in a first approximation for sufficiently small couplings and at sufficiently low energies. However, for sufficiently large couplings the system can spontaneously break a symmetry in the particle-hole channel $\langle\Psi^\dag M\Psi\rangle\neq 0$ or in the particle-particle channel $\langle\Psi^T N\Psi\rangle\neq 0$, where the Hermitian matrices $M,N$ describe the type of order that develops.

In this section we perform a one-loop RG calculation that allows us to explore the possible symmetry-breaking orders at strong coupling. We first derive RG equations that describe the flow of the coupling constants as the energy scale is lowered (Sec.~\ref{sec:oneloopRG}). We find a total of fifteen fixed points, including one stable (Gaussian) fixed point, four critical points, six bicritical points, and four tricritical points (Sec.~\ref{sec:FPs}). To explore the possible broken-symmetry states, we focus on the strong coupling regime and determine which asymptotic flows to strong coupling are stable (Sec.~\ref{sec:RGstroncoupling}). In Sec.~\ref{sec:susceptibilities}, we find the susceptibility that grows the fastest along a stable asymptotic flow to strong coupling, which determines the leading instability towards symmetry breaking.

\subsection{Lagrangian}
\label{sec:RGLagrangian}

We perform a one-loop Wilsonian RG calculation\cite{shankar1994} that consists in integrating out the high-energy fermionic modes in a thin frequency/momentum shell between $\Lambda/b$ and $\Lambda$, where $b=1+d\ell$ and $d\ell>0$ is an infinitesimal RG parameter. At the one-loop level, we find that there is no wave function or velocity renormalization, and for simplicity we set $v_\parallel=v_z=1$ in the Hamiltonian (\ref{fullH}). The RG calculation is simplest in the Lagrangian formalism, and the Lagrangian in Euclidean spacetime is
\begin{align}\label{Lag}
\mathcal{L}&=i\bar{\Psi}\slashed{\partial}\Psi+g_1(\bar{\Psi}\Gamma_4\Psi)^2
+g_2(\bar{\Psi}\Gamma_5\Psi)^2
+g_3(\bar{\Psi}\Gamma_{34}\Psi)^2\nonumber\\
&\hspace{4mm}+g_4[(\bar{\Psi}\Gamma_{14}\Psi)^2+
(\bar{\Psi}\Gamma_{24}\Psi)^2],
\end{align}
where $\slashed{\partial}=\gamma_\mu\partial_\mu$, the Dirac conjugate is $\bar{\Psi}=-i\Psi^\dag\gamma_0$, and we define a modified set of gamma matrices $\gamma_\mu$, $\mu=0,1,2,3$, by $\gamma_0=\Gamma_4$, $\gamma_1=-\Gamma_{14}$, $\gamma_2=-\Gamma_{24}$, $\gamma_3=-\Gamma_{34}$, and $\gamma_5=
\gamma_0\gamma_1\gamma_2\gamma_3
=\Gamma_{45}$. This facilitates the calculation of traces of products of gamma matrices. The matrices $\gamma_\mu$ satisfy the $SO(4)$ Clifford algebra, $\{\gamma_\mu,\gamma_\nu\}=2\delta_{\mu\nu}$, $\mu,\nu=0,1,2,3$, and $\{\gamma_5,\gamma_\mu\}=0$, $\gamma_5^2=1$. In terms of $\bar{\Psi}$, the quartic terms (\ref{X1X9}) are given by
\begin{align}\label{X1X9b}
X_1&=-(\bar{\Psi}\Gamma_4\Psi)^2,\nonumber\\
X_2&=(\bar{\Psi}\Gamma_3\Psi)^2
-(\bar{\Psi}\Gamma_{12}\Psi)^2,\nonumber\\
X_3&=-[(\bar{\Psi}\Gamma_0\Psi)^2
-(\bar{\Psi}\Gamma_{45}\Psi)^2],\nonumber\\
X_4&=-(\bar{\Psi}\Gamma_{35}\Psi)^2,\nonumber\\
X_5&=(\bar{\Psi}\Gamma_5\Psi)^2,\nonumber\\
X_6&=(\bar{\Psi}\Gamma_{34}\Psi)^2,\nonumber\\
X_7&=(\bar{\Psi}\Gamma_1\Psi)^2
+(\bar{\Psi}\Gamma_2\Psi)^2
-(\bar{\Psi}\Gamma_{13}\Psi)^2
-(\bar{\Psi}\Gamma_{23}\Psi)^2,\nonumber\\
X_8&=-[(\bar{\Psi}\Gamma_{15}\Psi)^2
+(\bar{\Psi}\Gamma_{25}\Psi)^2],\nonumber\\
X_9&=(\bar{\Psi}\Gamma_{14}\Psi)^2
+(\bar{\Psi}\Gamma_{24}\Psi)^2.
\end{align}
We have traded the couplings $\lambda_1,\lambda_5,\lambda_6,\lambda_9$ for $g_1,g_2,g_3,g_4$ and calculate the one-loop RG beta functions for the latter.

There is a subtlety in the RG procedure.\cite{herbut2009} Integrating out the high-energy fermionic modes will in general generate all the terms allowed by symmetry, i.e., all the quartic terms in Eq.~(\ref{X1X9b}). Naively, it would be impossible to obtain a closed set of equations for the couplings $g_1,g_2,g_3,g_4$. To avoid this, we use the Fierz identity to express the terms generated by integrating out the high-energy modes in terms of the chosen linearly independent couplings $X_1,X_5,X_6,X_9$. Using Gaussian elimination on the Fierz matrix (\ref{fierzF}), the equations to be used are
\begin{align}\label{GaussElim}
X_2&=-X_1+X_6-X_9,\nonumber\\
X_3&=X_1+2X_5+X_6+X_9,\nonumber\\
X_4&=-X_1-X_5-X_6,\nonumber\\
X_7&=-2X_1-2X_6,\nonumber\\
X_8&=-2X_1-2X_5-X_9.
\end{align}

\subsection{One-loop RG analysis}
\label{sec:oneloopRG}

\begin{figure}[t]
\begin{center}\includegraphics[width=\columnwidth]{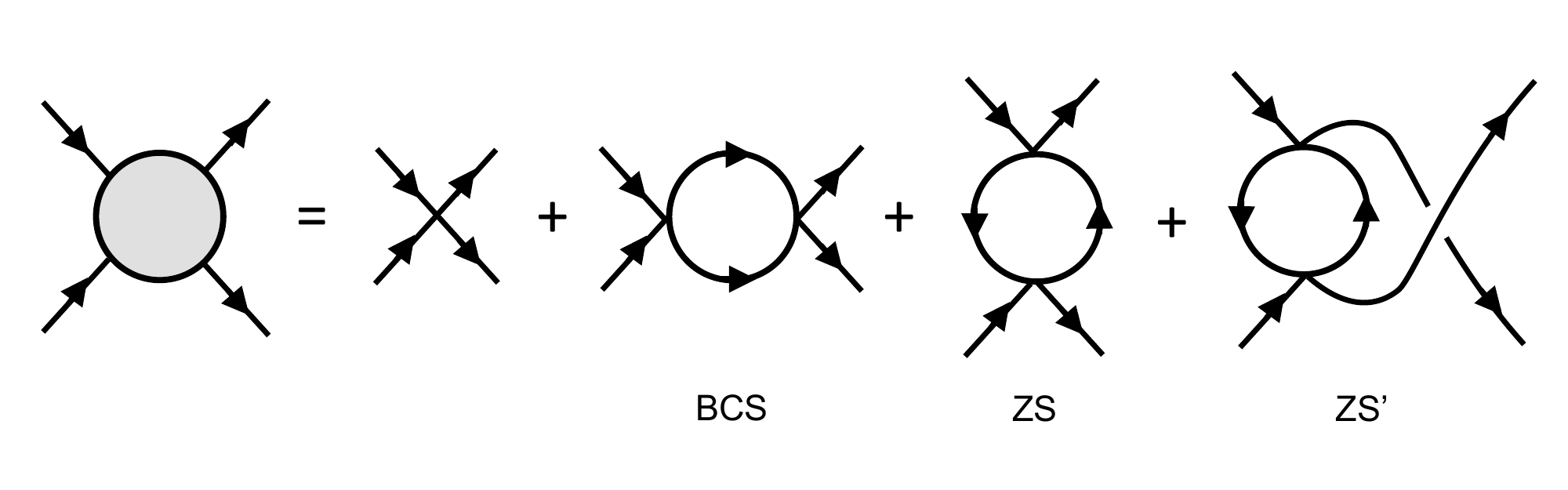}
\end{center}
\caption{One-loop renormalization of the four-fermion vertex.}
\label{fig:rg}
\end{figure}

Besides the tree-level term, three types of diagrams contribute to the one-loop RG beta function (Fig.~\ref{fig:rg}). If we write the interaction Lagrangian in Eq.~(\ref{Lag}) in the general form $\mathcal{L}_\textrm{int}=\sum_A g_A(\bar{\Psi}\Gamma_A\Psi)^2$, the contribution $\delta\mathcal{L}_<$ of the high-energy fermionic modes to the effective Lagrangian for the low-energy fermionic modes $\Psi^<,\bar{\Psi}^<$ consists of four terms,
\begin{align}
\delta\mathcal{L}_<=\delta\mathcal{L}_<^{(1)}
+\delta\mathcal{L}_<^{(2)}+\delta\mathcal{L}_<^{(3)}
+\delta\mathcal{L}_<^{(4)},
\end{align}
where
\begin{align}
\delta\mathcal{L}_<^{(1)}&=2\sum_{AB}g_Ag_B
\int_{\Lambda/b}^\Lambda\frac{d^4p}{(2\pi)^4}
\frac{\Tr\slashed{p}\Gamma_A\slashed{p}\Gamma_B}{(p^2)^2}
\nonumber\\
&\hspace{4mm}\times(\bar{\Psi}^<\Gamma_A\Psi^<)(\bar{\Psi}^<\Gamma_B\Psi^<),\\
\delta\mathcal{L}_<^{(2)}&=-2\sum_{A\neq B}g_Ag_B
\int_{\Lambda/b}^\Lambda\frac{d^4p}{(2\pi)^4}
\frac{1}{(p^2)^2}\nonumber\\
&\hspace{4mm}\times
(\bar{\Psi}^<\Gamma_A\slashed{p}\Gamma_B\Psi^<)(\bar{\Psi}^<\Gamma_B\slashed{p}\Gamma_A\Psi^<),\label{L2}\\
\delta\mathcal{L}_<^{(3)}&=-4\sum_{AB}g_Ag_B
\int_{\Lambda/b}^\Lambda\frac{d^4p}{(2\pi)^4}
\frac{1}{(p^2)^2}\nonumber\\
&\hspace{4mm}\times
(\bar{\Psi}^<\Gamma_A\Psi^<)(\bar{\Psi}^<\Gamma_B\slashed{p}
\Gamma_A\slashed{p}\Gamma_B\Psi^<),\label{L3}\\
\delta\mathcal{L}_<^{(4)}&=2\sum_{A\neq B}g_Ag_B
\int_{\Lambda/b}^\Lambda\frac{d^4p}{(2\pi)^4}
\frac{1}{(p^2)^2}\nonumber\\
&\hspace{4mm}\times
(\bar{\Psi}^<\Gamma_A\slashed{p}\Gamma_B\Psi^<)(\bar{\Psi}^<\Gamma_A\slashed{p}\Gamma_B\Psi^<),\label{L4}
\end{align}
where we define the frequency-momentum four-vector $p=(\omega,p_x,p_y,p_z)$, and $\slashed{p}=\gamma_\mu p_\mu$. Products of fermion bilinears in Eq.~(\ref{L2}), (\ref{L3}), and (\ref{L4}) are simplified by the use of the Fierz identity (\ref{fierzid}), for example,
\begin{align}
&(\bar{\Psi}^<\Gamma_A\slashed{p}\Gamma_B\Psi^<)(\bar{\Psi}^<\Gamma_B\slashed{p}\Gamma_A\Psi^<)\nonumber\\
&=-\frac{1}{16}(\Tr\Gamma_A\slashed{p}\Gamma_B\Gamma_C
\Gamma_B\slashed{p}\Gamma_A\Gamma_D)
(\bar{\Psi}^<\Gamma_C\Psi^<)(\bar{\Psi}^<\Gamma_D\Psi^<).
\end{align}
Performing the traces of products of gamma matrices and the momentum integrals, we obtain (dropping the superscripts $<$ on $\Psi^<,\bar{\Psi}^<$ for simplicity)
\begin{widetext}
\begin{align}
\delta\mathcal{L}_<^{(1)}=S_4\Lambda^2d\ell
\left\{-4g_1^2(\bar{\Psi}\Gamma_4\Psi)^2
+4g_2^2(\bar{\Psi}\Gamma_5\Psi)^2
-4g_3^2(\bar{\Psi}\Gamma_{34}\Psi)^2
-4g_4^2[(\bar{\Psi}\Gamma_{14}\Psi)^2
+(\bar{\Psi}\Gamma_{24}\Psi)^2]\right\},\nonumber
\end{align}
\begin{align}
\delta\mathcal{L}_<^{(2)}&=S_4\Lambda^2d\ell
\cdot\frac{1}{2}
\bigl\{2(g_1g_2+g_1g_3+g_2g_3+2g_1g_4+2g_2g_4
+2g_3g_4+g_4^2)[(\bar{\Psi}\Gamma_0\Psi)^2
-(\bar{\Psi}\Gamma_{45}\Psi)^2]\nonumber\\
&\hspace{20mm}+(g_1g_2+g_1g_3-g_2g_3+2g_1g_4-2g_2g_4-2g_3g_4-g_4^2)(\bar{\Psi}\Gamma_4\Psi)^2\nonumber\\
&\hspace{20mm}+(-g_1g_2-g_1g_3+g_2g_3-2g_1g_4+2g_2g_4+2g_3g_4+g_4^2)(\bar{\Psi}\Gamma_5\Psi)^2\nonumber\\
&\hspace{20mm}+(g_1g_2-g_1g_3+g_2g_3+g_4^2)[(\bar{\Psi}\Gamma_{14}\Psi)^2
+(\bar{\Psi}\Gamma_{24}\Psi)^2]\nonumber\\
&\hspace{20mm}+(-g_1g_2+g_1g_3-g_2g_3-g_4^2)
[(\bar{\Psi}\Gamma_{15}\Psi)^2
+(\bar{\Psi}\Gamma_{25}\Psi)^2]\nonumber\\
&\hspace{20mm}+(g_1g_2+g_1g_3-g_2g_3-2g_1g_4+2g_2g_4+2g_3g_4-g_4^2)(\bar{\Psi}\Gamma_{34}\Psi)^2\nonumber\\
&\hspace{20mm}+(-g_1g_2-g_1g_3+g_2g_3+2g_1g_4-2g_2g_4-2g_3g_4+g_4^2)(\bar{\Psi}\Gamma_{35}\Psi)^2\bigr\},\nonumber
\end{align}
\begin{align}
\delta\mathcal{L}_<^{(3)}&=S_4\Lambda^2d\ell\cdot 2
\bigl\{
g_1(g_1-g_2-g_3-2g_4)(\bar{\Psi}\Gamma_4\Psi)^2
+g_2(-g_2+g_1-g_3-2g_4)(\bar{\Psi}\Gamma_5\Psi)^2\nonumber\\
&\hspace{20mm}+g_3(g_3-g_1+g_2-2g_4)(\bar{\Psi}\Gamma_{34}\Psi)^2
+g_4(-g_1+g_2-g_3)[(\bar{\Psi}\Gamma_{14}\Psi)^2
+(\bar{\Psi}\Gamma_{24}\Psi)^2]\bigr\},\nonumber
\end{align}
\begin{align}
\delta\mathcal{L}_<^{(4)}&=S_4\Lambda^2d\ell
\cdot\left(-\frac{1}{2}\right)
\bigl\{2g_1g_2[(\bar{\Psi}\Gamma_0\Psi)^2
-(\bar{\Psi}\Gamma_{45}\Psi)^2]\nonumber\\
&\hspace{30mm}+2(-g_1g_4-g_2g_4+g_3g_4)
[(\bar{\Psi}\Gamma_1\Psi)^2
+(\bar{\Psi}\Gamma_2\Psi)^2
-(\bar{\Psi}\Gamma_{13}\Psi)^2
-(\bar{\Psi}\Gamma_{23}\Psi)^2]\nonumber\\
&\hspace{30mm}+2(-g_1g_3-g_2g_3+g_4^2)
[(\bar{\Psi}\Gamma_3\Psi)^2
-(\bar{\Psi}\Gamma_{12}\Psi)^2]\nonumber\\
&\hspace{30mm}+(g_1g_2-g_1g_3+g_2g_3-2g_1g_4+2g_2g_4-2g_3g_4-g_4^2)(\bar{\Psi}\Gamma_4\Psi)^2\nonumber\\
&\hspace{30mm}+(-g_1g_2+g_1g_3-g_2g_3+2g_1g_4-2g_2g_4+2g_3g_4+g_4^2)(\bar{\Psi}\Gamma_5\Psi)^2\nonumber\\
&\hspace{30mm}+(g_1g_2-g_1g_3+g_2g_3-2g_1g_4+2g_2g_4-2g_3g_4-g_4^2)
[(\bar{\Psi}\Gamma_{14}\Psi)^2
+(\bar{\Psi}\Gamma_{24}\Psi)^2]\nonumber\\
&\hspace{30mm}+(-g_1g_2+g_1g_3-g_2g_3+2g_1g_4-2g_2g_4+2g_3g_4+g_4^2)
[(\bar{\Psi}\Gamma_{15}\Psi)^2
+(\bar{\Psi}\Gamma_{25}\Psi)^2]\nonumber\\
&\hspace{30mm}+(g_1g_2-g_1g_3+g_2g_3-2g_1g_4+2g_2g_4-2g_3g_4-g_4^2)(\bar{\Psi}\Gamma_{34}\Psi)^2\nonumber\\
&\hspace{30mm}+(-g_1g_2+g_1g_3-g_2g_3+2g_1g_4-2g_2g_4+2g_3g_4+g_4^2)(\bar{\Psi}\Gamma_{35}\Psi)^2\bigr\},
\end{align}
\end{widetext}
where $S_4=1/8\pi^2$ is the surface area of the unit 3-sphere divided by $(2\pi)^4$. Using Eq.~(\ref{GaussElim}) to eliminate $X_2,X_3,X_4,X_7,X_8$ in favor of the linearly independent quartic terms $X_1,X_5,X_6,X_9$, we obtain
\begin{align}
\delta\mathcal{L}_<&=S_4\Lambda^2 d\ell
\bigl\{f_1(\{g\})(\bar{\Psi}\Gamma_4\Psi)^2
+f_2(\{g\})(\bar{\Psi}\Gamma_5\Psi)^2\nonumber\\
&\hspace{4mm}+f_3(\{g\})(\bar{\Psi}\Gamma_{34}\Psi)^2\nonumber\\
&\hspace{4mm}+f_4(\{g\})[(\bar{\Psi}\Gamma_{14}\Psi)^2
+(\bar{\Psi}\Gamma_{24}\Psi)^2]\bigr\},
\end{align}
where $f_1,f_2,f_3,f_4$ are quadratic polynomials in the coupling constants given by
\begin{align}\label{ffunctions}
f_1(\{g\})&=-2g_1^2-2g_1g_2+2g_1g_3+4g_1g_4+4g_3g_4+2g_4^2,\nonumber\\
f_2(\{g\})&=2g_1g_2+2g_2^2-4g_1g_3-2g_2g_3-8g_1g_4-4g_2g_4
\nonumber\\
&\hspace{4mm}-8g_3g_4-4g_4^2,\nonumber\\
f_3(\{g\})&=-2g_1g_3+2g_2g_3-2g_3^2-4g_1g_4-4g_2g_4
\nonumber\\
&-4g_3g_4-2g_4^2,\nonumber\\
f_4(\{g\})&=-2g_1g_3-2g_2g_3-4g_1g_4-4g_3g_4
-4g_4^2.
\end{align}
It is convenient to define dimensionless couplings $S_4\Lambda^2 g_i\rightarrow g_i$. Rescaling the high-energy cutoff $\Lambda$ and the fields, we obtain the four one-loop RG equations
\begin{align}\label{RGE}
\frac{dg_i}{d\ell}=-2g_i+f_i(\{g\}),\hspace{5mm}i=1,2,3,4,
\end{align}
which are the main result of this section. In the following section we analyze the fixed-point structure of these equations.

\subsection{Fixed points}
\label{sec:FPs}

The tree-level term $-2g_i$ in the RG equations (\ref{RGE}) implies that the Gaussian fixed point $(g_1,g_2,g_3,g_4)=(0,0,0,0)$ is stable. The other fixed points are given by nontrivial solutions of the system of four quadratic equations in four variables
\begin{align}
2g_i+f_i(\{g\})=0,\hspace{5mm}i=1,2,3,4,
\end{align}
which can be solved analytically. We find a total of fifteen fixed points (Table~\ref{tableone}). The Gaussian fixed point is the only stable one, besides which we find four critical points, six bicritical points, and four tricritical points.
\begin{table}[t]
\begin{tabular}{c||cccc|cccc|c}
\hline
FP & $g_1^*$ & $g_2^*$ & $g_3^*$ & $g_4^*$ & $y_1$ & $y_2$ & $y_3$ & $y_4$ & type \\
\hline\hline
1 & $0$ & $0$ & $0$ & $0$ & $-2$ & $-2$ & $-2$ & $-2$
& S \\ \hline
2 & $-1$ & $0$ & $0$ & $0$ & $-4$ & $4$ & $-2$ & $2$
& B \\ \hline
3 & $0$ & $1$ & $0$ & $0$ & $-4$ & $-4$ & $2$ & $2$
& B\\ \hline
4 & $0$ & $0$ & $-1$ & $0$ & $-4$ & $4$ & $-2$ & $2$
& B \\ \hline
5 & $-1$ & $1$ & $1$ & $0$ & $-4$ & $-4$ & $2$ & $2$
& B \\ \hline
6 & $-3$ & $4$ & $1$ & $1$ & $-10$ & $10$ & $10$ & $2$
& T \\ \hline
7 & $\frac{1}{3}$ & $\frac{2}{3}$ & $-\frac{2}{3}$ & $-\frac{2}{3}$
 & $-\frac{20}{3}$ & $\frac{20}{3}$ & $\frac{10}{3}$ & $2$
 & T \\
\hline
8 & $-\frac{4}{3}$ & $\frac{2}{3}$ & $1$ & $-\frac{2}{3}$
& $-\frac{20}{3}$ & $\frac{20}{3}$ & $\frac{10}{3}$ & $2$
& T \\
\hline
9 & $\frac{1}{8}$ & $-\frac{3}{8}$ & $-\frac{1}{4}$ & $-\frac{1}{4}$
& $-\frac{5}{2}$ & $-\frac{5}{2}$ & $-\frac{5}{2}$ & $2$
& C \\
\hline
10 & $-\frac{1}{7}$ & $\frac{3}{7}$ & $-\frac{3}{7}$ & $\frac{2}{7}$
& $-\frac{20}{7}$ & $-\frac{20}{7}$ & $2$ & $-\frac{10}{7}$
& C \\
\hline
11 & $-\frac{6}{7}$ & $\frac{3}{7}$ & $\frac{2}{7}$ & $\frac{2}{7}$
& $-\frac{20}{7}$ & $-\frac{20}{7}$ & $2$ & $-\frac{10}{7}$
& C \\
\hline
12 & $\frac{\sqrt{5}-2}{4}$ & $\frac{1}{4}$ & $\frac{\sqrt{5}-1}{4}$ & $-\frac{\sqrt{5}+1}{4}$
& $-5$ & $5$ & $-\sqrt{5}$ & $2$
& B \\
\hline
13 & $-\frac{3-\sqrt{5}}{6}$ & $\frac{2}{3}$ & $\frac{\sqrt{5}+1}{6}$ & $-\frac{\sqrt{5}-1}{6}$
& $-\frac{10}{3}$ & $-\frac{10}{3}$ & $2$ & $-\frac{2\sqrt{5}}{3}$
& C\\
\hline
14 & $-\frac{\sqrt{5}+2}{4}$ & $\frac{1}{4}$ & $-\frac{\sqrt{5}+1}{4}$ & $\frac{\sqrt{5}-1}{4}$
& $-5$ & $5$ & $\sqrt{5}$ & $2$
& T \\
\hline
15 & $-\frac{3+\sqrt{5}}{6}$ & $\frac{2}{3}$ & $-\frac{\sqrt{5}-1}{6}$ & $\frac{\sqrt{5}+1}{6}$
& $-\frac{10}{3}$ & $-\frac{10}{3}$ & $2$ & $\frac{2\sqrt{5}}{3}$
& B\\
\hline
\end{tabular}
\caption{RG fixed points (FP) at one-loop, with eigenvalues $y_i$ of the linearized RG equations and type of fixed point (S: stable, C: critical, B: bicritical, T: tricritical).}
\label{tableone}
\end{table}

\subsection{RG flows at strong coupling}
\label{sec:RGstroncoupling}

We have shown that the noninteracting fixed point is the only finite-coupling, stable fixed point of the one-loop RG equations. Thus under RG with generic initial conditions, the Weyl semimetal must either flow to the Gaussian fixed point, or to strong coupling. We now analyze the flows to strong coupling that represent the potential instabilities of the Weyl semimetal. Different instabilities are represented by different ``fixed trajectories'', i.e., different directions in which we can flow to strong coupling in the four-dimensional space of coupling constants $g_1,g_2,g_3,g_4$.  We will show that there is a single stable fixed trajectory towards strong coupling. This implies that any integration of the one-loop RG equations that flows to strong coupling must do so in the direction of the unique stable fixed trajectory. Thus, the one-loop RG analysis predicts a unique instability.

When analyzing the asymptotic flow to strong coupling, we can neglect the tree-level terms in the RG beta functions. Thus we obtain asymptotic one-loop RG equations of the form
\begin{align}
\frac{dg_i}{d\ell}=f_i(\{g\}),\hspace{5mm}i=1,2,3,4,
\end{align}
where the functions $f_i$ are given in Eq.~(\ref{ffunctions}). These equations have the scaling solution
\begin{align}\label{gscaling}
g_i(\ell) = \frac{G_i}{\ell_c - \ell},
\end{align}
where $G_1,G_2,G_3,G_4$ are constants. Substituting the scaling solution into the above differential equations yields a set of algebraic equations
\begin{align}
G_i=f_i(\{G\}),\hspace{5mm}i=1,2,3,4.
\end{align}
These algebraic equations specify the asymptotic ratios of the various couplings as the system flows to strong coupling. There are fourteen nontrivial solutions to the above set of equations (Table~\ref{flows}). These non-trivial solutions are the directions in parameter space along which the system can flow to strong coupling. However, not all the solutions are stable. To investigate the (linear) stability of a solution, one must consider small perturbations $\delta_i$ from the fixed trajectory. Linearizing the flow equations in small perturbations $\delta_i$ about the fixed trajectory, we obtain the linearized flow equations $\frac{d\delta_i}{d\ell} = M_{ij} \delta_j$, where
\begin{widetext}
\begin{eqnarray}
M &=& \left( \begin{array}{cccc} - 2 G_1 - G_2 + G_3 + 2 G_4 & - G_1 & G_1 + 2 G_4 & 2 G_1 + 2 G_3 + 2 G_4 \\
G_2 - 2 G_3 - 4 G_4 & G_1 + 2 G_2 - G_3 - 2 G_4 & - 2 G_1 - G_2 - 4 G_4 & - 4 G_1 - 2 G_2 - 4 G_3 - 4 G_4 \\
- G_3 - 2 G_4 & G_3 - 2 G_4 & -G_1 + G_2 - 2 G_3 - 2 G_4 & - 2 G_1 - 2 G_2 - 2 G_3 - 2 G_4 \\
- G_3 - 2 G_4 & - G_3 & - G_1 - G_2 - 2 G_4 & - 2 G_1 - 2 G_2 - 4 G_4 \end{array}\right).\nonumber
\end{eqnarray}
\end{widetext}
The stability matrix $M$ necessarily has one positive eigenvalue, corresponding to flow along the fixed trajectory. One should thus project onto the subspace orthogonal to the fixed trajectory by acting with $\delta_{ij} - P_{ij}$, where $P_{ij} = \frac{1}{G_1^2 + G_2^2 + G_3^2 + G_4^2} G_i G_j$. To determine the stability of a particular fixed trajectory, one should look at eigenvalues of the projected stability matrix $M(I-P)$ where $I$ is the $4\times 4$ identity matrix. If this matrix has any positive eigenvalues, the fixed trajectory is unstable. 

Performing a stability analysis about the fourteen possible fixed trajectories, we find that only one of them is stable, i.e., has a projected stability matrix with strictly negative eigenvalues. This is the trajectory with $(G_1, G_2, G_3, G_4) = (\frac{1}{16}, -\frac{3}{16}, -\frac{1}{8}, -\frac{1}{8})$. Since there is a unique stable fixed trajectory, there is a unique flow to strong coupling. Thus, within the one-loop RG, either the system flows to the Gaussian fixed point, or it flows to strong coupling, with the various couplings in the ratio $(G_1 : G_2 : G_3 : G_4) = (1, -3, -2, -2)$ and with signs $G_1 > 0, G_2 < 0, G_3 = G_4 < 0$. This is the {\it only} possible flow to strong coupling that can be obtained starting from generic initial conditions and using the one-loop RG equations. For future reference, the Hamiltonian for the fixed trajectory (which has $g_3=g_4$) takes the form
\begin{align}\label{HFixedTrajectory}
H &= H_0 + \int d^3r\,\biggl(g_A (\Psi^{\dag}\Psi)^2 + g_B (\Psi^{\dag} \Gamma_{45} \Psi)^2\nonumber\\
& \hspace{4mm}+ g_C \sum_{i=1}^3 (\Psi^{\dag} \Gamma_{i} \Psi)^2\biggr), 
\end{align}
where $g_A = -g_1$, $g_B =  g_2$, and $g_C =  g_3$, and the fixed trajectory has $(g_A, g_B, g_C) = -\frac{1}{16}(1,3,2)$.  Unlike the general Hamiltonian (\ref{fullH}), the Hamiltonian for the fixed trajectory (\ref{HFixedTrajectory}) has an emergent $SO(3)$ rotation symmetry (see Appendix~\ref{app:SymmSCOP}). The emergence of larger symmetries near certain fixed points was discussed previously in the analogous context of fermions with short-range interactions in graphene.\cite{roy2011}

\begin{table}[t]
\begin{tabular}{cccc|c}
\hline
$G_1$ & $G_2$ & $G_3$ & $G_4$ & stability\\
\hline\hline
$\frac{1}{6}$ & $\frac{1}{3}$ & $-\frac{1}{3}$ & $-\frac{1}{3}$ & U\\
\hline
$-\frac{2}{3}$ & $\frac{1}{3}$ & $\frac{1}{2}$ & $-\frac{1}{3}$ & U\\
\hline
$\frac{1}{16}$ & $-\frac{3}{16}$ & $-\frac{1}{8}$ & $-\frac{1}{8}$& S\\
\hline
$0$ & $0$ & $-\frac{1}{2}$ & $0$ & U\\
\hline
$-\frac{1}{2} $ & $0$ & $0$ & $0$ & U\\
\hline
$0$ & $\frac{1}{2}$ & $0$ & $0$ & U\\
\hline
$-\frac{1}{2}$ & $\frac{1}{2}$ & $\frac{1}{2}$ & $0$ & U\\
\hline
$-\frac{1}{14}$ & $\frac{3}{14}$ & $-\frac{3}{14}$ & $\frac{1}{7}$ & U\\
\hline
$-\frac{3}{7}$ & $\frac{3}{14}$ & $\frac{1}{7}$ & $\frac{1}{7}$ & U\\
\hline
$-\frac{3}{2}$ & $2$ & $\frac{1}{2}$ & $\frac{1}{2}$ & U\\
\hline
$\frac{7 - 3 \sqrt5}{8(\sqrt5 - 1)}$ & $\frac{1}{8}$ & $\frac{\sqrt5 - 1}{8}$ & $-\frac{\sqrt5 + 1}{8}$ & U\\
\hline
$-\frac{7 + 3 \sqrt5}{8(\sqrt5 + 1)}$ & $\frac{1}{8}$ & $-\frac{\sqrt5 + 1}{8}$ & $\frac{\sqrt5 - 1}{8}$ & U\\
\hline
$-\frac{\sqrt5 - 1}{6(\sqrt5 + 1)}$ & $\frac{1}{3}$ & $\frac{\sqrt5 + 1}{12}$ & $-\frac{\sqrt5 - 1}{12}$ & U\\
\hline
$-\frac{\sqrt5 + 1}{6(\sqrt5 - 1)}$ & $\frac{1}{3}$ & $-\frac{\sqrt5 - 1}{12}$ & $\frac{\sqrt5 + 1}{12}$ & U\\
\hline
\end{tabular}
\caption{\label{flows} Fixed-point trajectories for flows to strong coupling, and their stability (S: stable, U: unstable). There is a unique stable fixed trajectory, and thus a unique stable flow to strong coupling. }
\end{table}

\section{Order parameters and instabilities}
\label{sec:susceptibilities}

Having identified the direction in which the system flows to strong coupling, we focus on what kind of ordering can be expected at strong coupling. To this end, we consider the susceptibility towards developing an expectation value for all possible momentum-independent fermion bilinears, in both the particle-hole and particle-particle channels. The largest susceptibility indicates the leading instability along the fixed trajectory. 

\subsection{Particle-hole channels}

\begin{figure}[t]
\begin{center}\includegraphics[width=\columnwidth]{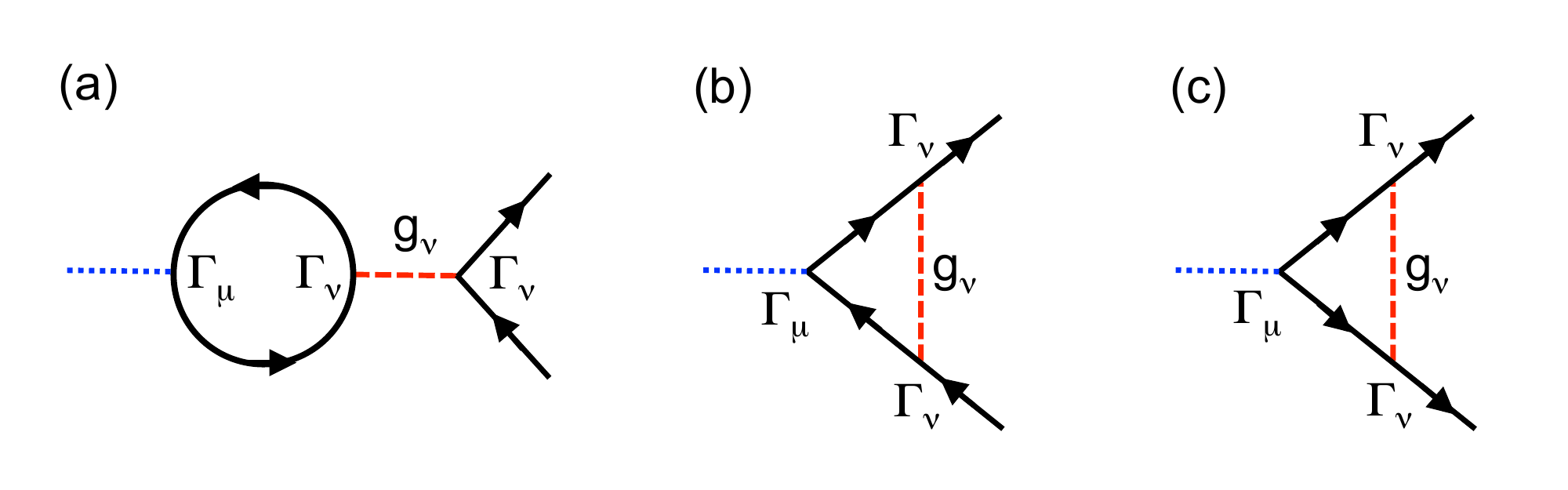}
\end{center}
\caption{A test vertex in the particle-hole channel with structure $\Gamma_\mu$ renormalizes through the diagrams (a) and (b). A test vertex in the particle-particle channel renormalizes through the diagram (c). We are using a diagrammatic code wherein dotted blue lines indicate test vertices, dashed red lines indicate interactions, and solid black lines indicate fermion Green's functions (color online).}
\label{vertexrenormalizations}
\end{figure}

We illustrate the procedure by adding to the Lagrangian a test vertex in the particle-hole channel of the form $\Delta_\mu^\textrm{ph}\Psi^{\dag}\Gamma_\mu\Psi$, where $\Gamma_\mu$ is one of the fifteen gamma matrices $\Gamma_a,\Gamma_{ab}$, $a,b=1,\ldots,5$, $a<b$ ($\Gamma_0$ simply corresponds to a global shift of the chemical potential). This vertex renormalizes through the diagrams shown in Fig.~\ref{vertexrenormalizations}(a),(b), and also has a tree-level scaling dimension of $+1$. Thus, the RG flow equation for the vertex $\Delta_\mu^\textrm{ph}$ takes the form

\begin{equation}
\frac{d \ln{\Delta_\mu^\textrm{ph}}}{ d\ell }
= 1 + \sum_{i} A_{\mu i}^\textrm{ph} g_i,
\end{equation}
where the coefficients $A_{\mu i}$ remain to be determined. Substituting the strong-coupling scaling form (\ref{gscaling}) into the above equation and solving yields
\begin{equation}
\chi_\mu^\textrm{ph}(\ell) = \frac{\Delta_\mu^\textrm{ph}(\ell)}{\Delta_\mu^\textrm{ph}(0)} = (\ell_c-\ell)^{-f_\mu^\textrm{ph}}, \,
 f_\mu^\textrm{ph} = \sum_{i} A_{\mu i}^\textrm{ph} G_i,
\end{equation}
Where we have defined the susceptibility $\chi_\mu^\textrm{ph}$. Thus, if $f_\mu^\textrm{ph} > 0$ there is a divergence in the susceptibility indicating an instability to ordering in this channel, with the largest divergence occurring in the channel with largest exponent $f_\mu^\textrm{ph}$. We now derive the coefficients $A_{\mu i}^\textrm{ph}$ and hence the $f_\mu^\textrm{ph}$.

For ordering in the channel $\Psi^{\dag} \Gamma_{\mu} \Psi$, the diagram in Fig.~\ref{vertexrenormalizations}(a) gives a contribution
\begin{align}
 \delta_{(a)} \ln \Delta_{\mu}^\textrm{ph} &=  g_A \Tr \Gamma_{\mu} G(\varepsilon, \b{k}) G(\varepsilon, \b{k})\nonumber\\
 &\hspace{4mm} + g_B \Tr \Gamma_{\mu} G(\varepsilon, \b{k}) \Gamma_{45} G(\varepsilon, \b{k})\nonumber\\
 &\hspace{4mm}  + g_C  \Tr \Gamma_{\mu} G(\varepsilon, \b{k}) (\Gamma_1 + \Gamma_2 + \Gamma_3) G(\varepsilon, \b{k}),
\end{align}
where the fermion Green's function is
\begin{align}
G(\varepsilon, \b{k}) = \left(-i \varepsilon +  \sum_{i=1}^3 v_i k_i \Gamma_i \right)^{-1} = \frac{i\varepsilon+  \sum_{i=1}^3 v_i k_i \Gamma_i}{\varepsilon^2 + v_{\parallel}^2 \b{k}_{\parallel}^2 + v^2_{z} k_z^2},
\end{align}  
and the traces are taken over spin and valley indices, and also indicate integration over $\varepsilon$ and $\b{k}$. The minus sign coming from the fermion loop has been cancelled by the minus sign associated with going up one order in perturbation theory. The integral over $\varepsilon$ is over the entire real line $-\infty < \varepsilon < \infty$, whereas the integration over $\b{k}$ is over an ellipsoidal shell of states with energy $e^{\ell-d\ell} < \sqrt{\sum_{i=1}^3 v^2_i k^2_i} < e^\ell$. It is convenient at this point to rescale $v_i k_i \rightarrow k_i$. This rescaling makes the Green's function isotropic, and allows us to take the $\b{k}$ integration over a spherical shell $e^{\ell-d\ell} < k < e^{\ell}$ which is easier to work with than an ellipsoidal shell. 
  
  The first trace vanishes for any $\Gamma_{\mu}$. The second and third traces vanish unless $\Gamma_{\mu} = \Gamma_{1}, \Gamma_2, \Gamma_3, \Gamma_{45}$. For $\Gamma_{\mu} = \Gamma_{1,2,3}$, only the third trace is nonzero. This gives
  \begin{align}
&\delta_{(a)} \ln \Delta_{1,2,3}^\textrm{ph}\nonumber\\
&= g_C \Tr \frac{ \Gamma_{1,2,3} (i\varepsilon+  \sum_i  k_i \Gamma_i) (\sum_i\Gamma_i) (i\varepsilon+ \sum_i k_i \Gamma_i)}{(\varepsilon^2 + k^2)^2}
\nonumber\\
  &= g_C \Tr \frac{-\varepsilon^2 + \sum_{i,j} k_i k_j \Gamma_{1,2,3} \Gamma_i ( \Gamma_1 + \Gamma_2 + \Gamma_3) \Gamma_j  }{(\varepsilon^2 + k^2 )^2}\nonumber\\
    &= g_C \Tr \frac{-\varepsilon^2 + \sum_i  k_i^2 \Gamma_{1,2,3} \Gamma_i ( \Gamma_1 + \Gamma_2 + \Gamma_3) \Gamma_i  }{(\varepsilon^2 + k^2 )^2}\nonumber\\
  &=  - g_C \int\frac{d\varepsilon}{2\pi}\int'\frac{d^3k}{(2\pi)^3} \frac{\varepsilon^2 + \frac13 k^2 }{(\varepsilon^2 + k^2)^2},
  \end{align}
where we have made use of the fact that any term odd in $\varepsilon$ vanishes upon integration over $\varepsilon$, and any term odd in $k_i$ vanishes upon integration over $k_i$. We have also used the relations $\int' d^3 k (k_1^2 - k_2^2 - k_3^2) f(k^2) = -\frac{1}{3} \int' d^3 k k^2 f(k^2)$ when the primed integral sign denotes integration over the spherical momentum shell, $\Gamma_i^2 = 1$, and $\Gamma_1 \Gamma_2 = - \Gamma_2 \Gamma_1$. We obtain
  \begin{equation}
  \delta_{(a)} \ln \Delta_{1,2,3}^\textrm{ph} = -\frac{4}{6\pi^2} g_C e^{2\ell} d\ell.
  \end{equation}
For ordering in the channel $\Psi^{\dag} \Gamma_{45} \Psi$, the diagram in Fig.~\ref{vertexrenormalizations}(a) gives a correction   %
  \begin{eqnarray}
  \delta_{(a)} \ln \Delta_{45}^\textrm{ph} &=& g_C \Tr \frac{- \varepsilon^2 + \sum_{i=1}^3 k_i^2 \Gamma_{45} \Gamma_i \Gamma_{45} \Gamma_i}{(\varepsilon^2 + k^2)^2}\nonumber\\
  &=& g_C \int\frac{d\varepsilon}{2\pi}\int'\frac{d^3k}{(2\pi)^3} \frac{- \varepsilon^2 + k^2}{(\varepsilon^2 + k^2)^2}\nonumber\\
  &=& 0,
  \end{eqnarray}
where we have made use of the identity $\Gamma_{1,2,3} \Gamma_{45} = \Gamma_{45} \Gamma_{1,2,3}$. For all other ordering channels the trace over the bubble in Fig.~\ref{vertexrenormalizations}(a) is trivially zero. Thus, the diagram in Fig.~\ref{vertexrenormalizations}(a) contributes only to the susceptibility in the particle-hole channel with structure $\Gamma_{1,2,3}$. 
  
The diagram in Fig.~\ref{vertexrenormalizations}(b) contributes to ordering in a particle-hole channel with structure $\Gamma_{\mu}$ as 
\begin{align}
\delta_{(b)}  \ln \Delta_{\mu}^\textrm{ph} &= - \int\frac{d\varepsilon}{2\pi}\int'\frac{d^3k}{(2\pi)^3}  \Bigl( g_A G(\varepsilon, \b{k}) \Gamma_{\mu} G(\varepsilon, \b{k})\nonumber\\
&\hspace{4mm} + g_B \Gamma_{45} G(\varepsilon, \b{k}) \Gamma_{\mu} G(\varepsilon, \b{k}) \Gamma_{45}\nonumber\\
&\hspace{4mm} + g_C \sum_{i=1}^3 \Gamma_i G(\varepsilon, \b{k}) \Gamma_{\mu} G(\varepsilon, \b{k}) \Gamma_i \Bigr),
  \end{align}
where there is a relative minus sign compared to the diagram in Fig.~\ref{vertexrenormalizations}(a) because of the lack of a fermion loop, and there is no trace over spin/valley indices, again because we do not have a fermion loop in this diagram. We can now use the various (anti)commutation relations for the gamma matrices to move these matrices in the above expression all the way over to the right. For $\mu = 1,2,3,$ we have
  \begin{align}
&\delta_{(b)} \ln \Delta_{1,2,3}^\textrm{ph}\nonumber\\
& = \int\frac{d\varepsilon}{2\pi}\int'\frac{d^3k}{(2\pi)^3} \frac{(g_A+g_B - g_C) (\varepsilon^2 + \frac13 k^2)  }{(\varepsilon^2 + k^2)^2}\nonumber\\
& = \frac{1}{6\pi^2} (g_A + g_B -  g_C) e^{2\ell} d\ell.
  \end{align}
Combining the renormalization from the diagrams in Fig.~\ref{vertexrenormalizations}(a) and (b), we obtain 
    \begin{equation}
  \delta \ln \Delta_{1,2,3}^\textrm{ph} = \frac{1}{6\pi^2} (g_A + g_B - 5 g_C) e^{2\ell} d\ell,
  \end{equation}
hence
\begin{align}
f_{1,2,3}^\textrm{ph}  =\frac{e^{2\ell_c}}{6\pi^2} (g_A + g_B - 5 g_C)=  \frac{e^{2\ell_c}}{16\pi^2} > 0,
\end{align}
indicating a triply degenerate instability to ordering in this channel, with coefficient $\frac{e^{2\ell_c}}{16\pi^2}$. For $\mu = 45$, a similar argument gives 
  \begin{align}
  \delta_{(b)} \ln \Delta_{45}^\textrm{ph} &= \int\frac{d\varepsilon}{2\pi}\int'\frac{d^3k}{(2\pi)^3} \frac{(g_A +g_B + 3 g_C)(\varepsilon^2 - k^2) }{(\varepsilon^2 + k^2)^2}\nonumber\\
  & = 0, 
  \end{align}
thus a test vertex in this channel is not renormalized by either diagram. There is no instability in this channel. 
  
For $\mu = 4, 5$, pushing gamma matrices to the right gives
\begin{align}
\delta_{(b)} \ln \Delta_{4,5}^\textrm{ph}&= \int\frac{d\varepsilon}{2\pi}\int'\frac{d^3k}{(2\pi)^3} \frac{(g_A - g_B - 3 g_C )(\varepsilon^2 + k^2)  }{(\varepsilon^2 + k^2)^2}\nonumber\\
&   = \frac{1}{4\pi^2} (g_A - g_B -  3 g_C) e^{2\ell} d\ell,
  \end{align}
hence
\begin{align}
f_{4,5}^\textrm{ph} = \frac{e^{2\ell_c}}{4 \pi^2} (g_A - g_B - 3 g_C) = \frac{e^{2\ell_c}}{8 \pi^2} > 0,
\end{align}
indicating a doubly degenerate instability in this channel, with larger coefficient than the instability in the $\Gamma_{1,2,3}$ channels. For $\mu = 12, 23, 13$, pushing gamma matrices to the right gives
  \begin{align}
&\delta_{(b)} \ln \Delta_{12, 23, 13}^\textrm{ph}\nonumber\\
& = \int\frac{d\varepsilon}{2\pi}\int'\frac{d^3k}{(2\pi)^3} \frac{(g_A + g_B - g_C )(\varepsilon^2 +\frac13 k^2)  }{(\varepsilon^2 + k^2)^2}\nonumber\\
&   = \frac{1}{6\pi^2} (g_A + g_B -  g_C) e^{2\ell} d\ell < 0,
  \end{align}
indicating no instability in this channel.  Finally, for $\mu = 14,24,34,15,25,35$ we obtain 
      \begin{align}
& \delta_{(b)} \ln \Delta_{14,24,34,15,25,35}^\textrm{ph}\nonumber\\
& = \int\frac{d\varepsilon}{2\pi}\int'\frac{d^3k}{(2\pi)^3} \frac{(g_A-g_B+g_C)(\varepsilon^2 - \frac13 k^2)}{(\varepsilon^2 + k^2)^2}\nonumber\\
& =  \frac{1}{12 \pi^2 }(g_A - g_B + g_C) e^{2\ell} d\ell = 0,
  \end{align}
indicating no instability in this channel. 
  
  Thus, there are instabilities in the particle-hole channel towards developing an expectation value for $\Psi^{\dag} \Gamma_{\mu} \Psi$, with $\mu = 1,2,3,4,5$ only. The leading instability is a doubly degenerate instability to ordering in a channel with $\mu = 4, 5$. As will be seen in Sec.~\ref{sec:LeadingInstability}, this type of order would gap out the Weyl points, and corresponds to SDW order at momentum $2Q$ in the $z$ direction, with an associated complex order parameter $M$ with $\Re M = \langle \Psi^{\dag} \Gamma_4 \Psi \rangle $ and $\Im M = \langle \Psi^{\dag} \Gamma_5 \Psi \rangle$.  There is also a subleading instability to ordering with $\mu = 1,2,3$ that corresponds to a type of intra-node ferromagnetism that simply shifts the position of the Weyl nodes, but this will likely be preempted by the leading instability, which destroys the Weyl nodes. 
  
  
  \subsection{Particle-particle channels}
  \label{sec: pp}
  
We now consider the particle-particle channels. These renormalize according to Fig.~\ref{vertexrenormalizations}(c). The possible test pairing vertices added to the Lagrangian are of the form $\Delta_\mu^\textrm{pp}\Psi^T\Gamma_{\mu}\Psi+\mathrm{h.c.}$, with the additional constraint from Fermi statistics that $\Gamma_{\mu}$ must be an antisymmetric matrix. This restricts us to $\mu = 2, 5, 13, 14, 25, 34$. We now obtain the vertex renormalization for each of these. From Fig.~\ref{vertexrenormalizations}(c) we obtain 
\begin{align}
\delta_{(b)} \ln \Delta^\textrm{pp}_{\mu} &= - \int\frac{d\varepsilon}{2\pi}\int'\frac{d^3k}{(2\pi)^3}  \Bigl( g_A G^T(-\varepsilon, -\b{k}) \Gamma_{\mu} G(\varepsilon, \b{k})\nonumber\\
&\hspace{4mm} + g_B \Gamma_{45} G^T(-\varepsilon, -\b{k}) \Gamma_{\mu} G(\varepsilon, \b{k}) \Gamma^T_{45}\nonumber\\
&\hspace{4mm} + g_C \sum_{i=1}^3 \Gamma_i G^T(-\varepsilon, -\b{k}) \Gamma_{\mu} G(\varepsilon, \b{k}) \Gamma^T_i \Bigr).
\end{align}
Again, there is no trace over spin/valley indices because there is no fermion loop. We can further simplify by noting that $G^T(-\varepsilon, -k_x, -k_y, -k_z) = - G(\varepsilon, k_x, - k_y, k_z)$, and also by noting that $\Gamma_{1,3,45}$ are symmetric matrices whereas $\Gamma_2$ is antisymmetric. We now check each of the channels in turn.  It is convenient to introduce the (modified) slashed notation $\slashed{k} = \sum_{i=1}^3 k_i \Gamma_i$ to be used throughout Sec.~\ref{sec: pp}. This is a slightly different slashed notation to the one introduced earlier (which involved the $\gamma$ matrices rather than the $\Gamma$ matrices), but it is the most convenient for our present purposes.
  %

  For particle-particle pairing in the $\Gamma_2$ channel, we use $\Gamma_2 (k_1 \Gamma_1 - k_2 \Gamma_2 + k_3 \Gamma_3 ) = - \slashed{k} \Gamma_2$ and thus $(i \varepsilon + \slashed{k})\Gamma_2  (k_1 \Gamma_1 - k_2 \Gamma_2 + k_3 \Gamma_3 ) = (i\varepsilon + \slashed{k})(i\varepsilon - \slashed{k}) \Gamma_2 = -(\varepsilon^2 + k^2) \Gamma_2$. Using also the commutation relations $[\Gamma_{45}, \Gamma_2] = 0 $ and $ \{ \Gamma_{1,3}, \Gamma_2 \} = 0$, we obtain
  \begin{align}
  f_2^\textrm{pp} = - \int\frac{d\varepsilon}{2\pi}\int'\frac{d^3k}{(2\pi)^3} \frac{(g_A + g_B - 3 g_C )(\varepsilon^2 +k^2)  }{(\varepsilon^2 + k^2)^2} < 0, 
  \end{align}
indicating no instability in this channel.    For $\mu = 13$, we obtain $f_{13}^\textrm{pp} = f_{2}^\textrm{pp}$. This follows because $\Gamma_2 (k_1 \Gamma_1 - k_2 \Gamma_2 + k_3 \Gamma_3 ) = - \slashed{k} \Gamma_2$ and $\Gamma_{13} (k_1 \Gamma_1 - k_2 \Gamma_2 + k_3 \Gamma_3 ) = - \slashed{k} \Gamma_{13}$, and $\Gamma_2$ and $\Gamma_{13}$ have the same commutation relations with $\Gamma_{1,2,3,45}$. 

In the $\Gamma_5$ channel we have $(i \varepsilon + \slashed{k})\Gamma_5  (k_1 \Gamma_1 - k_2 \Gamma_2 + k_3 \Gamma_3 ) = (i\varepsilon + \slashed{k})(i\varepsilon - k_1 \Gamma_1 + k_2 \Gamma_2 - k_3 \Gamma_3) \Gamma_5 = -(\varepsilon^2 +\frac13 k^2) \Gamma_5$. We have also $\{ \Gamma_{45}, \Gamma_5\} = 0$ and $\{ \Gamma_{1,2,3}, \Gamma_{5} \} = 0$, hence
  \begin{align}
  f_5^\textrm{pp} = - \int\frac{d\varepsilon}{2\pi}\int'\frac{d^3k}{(2\pi)^3} \frac{(g_A - g_B -  g_C )(\varepsilon^2 + \frac13  k^2)  }{(\varepsilon^2 + k^2)^2} < 0,
  \end{align}
indicating no instability in this channel.

  For the $\Gamma_{14}$ channel we have $(i \varepsilon + \slashed{k})\Gamma_{14}  (k_1 \Gamma_1 - k_2 \Gamma_2 + k_3 \Gamma_3 ) = (i\varepsilon + \slashed{k})(i\varepsilon - k_1 \Gamma_1 - k_2 \Gamma_2 + k_3 \Gamma_3) \Gamma_{14} = -(\varepsilon^2 +\frac13 k^2) \Gamma_{14}$, just like in the $\Gamma_{5}$ channel, and likewise in the $\Gamma_{23}$ channel. Also, $\{ \Gamma_{45}, \Gamma_{14, 34, 5} \} = 0$, thus $g_A$ and $g_B$ affect the $\Gamma_{14, 23}$ channels in the same way that they affect the $\Gamma_5$ channel. Meanwhile, keeping track of the transposition when evaluating the $g_C$ correction term,  %
   \begin{equation}
  f_{14, 23}^\textrm{pp} = - \int\frac{d\varepsilon}{2\pi}\int'\frac{d^3k}{(2\pi)^3} \frac{(g_A - g_B - g_C )(\varepsilon^2 + \frac13  k^2)  }{(\varepsilon^2 + k^2)^2}  < 0,
  \end{equation} 
again indicating no instability. The $\Gamma_{34}$ channel is degenerate with the $\Gamma_5$ channel, a consequence of $SO(2)$ rotation symmetry (see Appendix~\ref{app:SymmSCOP}). More unexpectedly, it is also degenerate with the $\Gamma_{14}$ channel. This follows because our model has an $SO(3)$ rotation invariance when $g_3=g_4$, which is the case along the fixed trajectory, and thus we can rotate $\Gamma_5$ into $\Gamma_{14}$ by acting with the matrix $\Gamma_{23}$ which corresponds to a rotation about the $x$ axis (see Appendix \ref{app:SymmSCOP}). Finally, for $\Gamma_{25}$ we have $(i \varepsilon + \slashed{k}) \Gamma_{25} (i \varepsilon + k_1 \Gamma_1 - k_2 \Gamma_2 + k_3 \Gamma_3) = (i \varepsilon + \slashed{k})(i\varepsilon + \slashed{k}) \Gamma_{25} = (-\varepsilon^2 + k^2) \Gamma_{25}$, which vanishes upon integration over energies. Thus there is no instability in the particle-particle channel. 
  
  
\section{Leading instability: spin-density wave ground state}
\label{sec:LeadingInstability}
  
\subsection{Mean-field Hamiltonian}
  
  The leading instability is in the particle-hole channel and is doubly degenerate, and the corresponding ordered state is described by the mean-field Hamiltonian 
  \begin{equation}\label{HMF}
  H_\textrm{MF} = H_0 + \int d^3r\left( \Delta_4 \Psi^{\dag} \Gamma_4 \Psi + \Delta_5 \Psi^{\dag} \Gamma_5 \Psi\right),
  \end{equation}
with $\Delta_4,\Delta_5$ real. The single-particle spectrum is fully gapped, $E_{\pm}(\b{p}) = \pm \sqrt{\b{p}^2 + \Delta_4^2 + \Delta_5^2}$. Using $\Psi^\dag=i\bar{\Psi}\gamma_0$ (see Sec.~\ref{sec:RGLagrangian} for the definition of the modified gamma matrices $\gamma_\mu$), the corresponding Euclidean Lagrangian is
 \begin{equation}
\mathcal{L}_\textrm{MF} = i\bar{\Psi}\gamma_\mu\partial_\mu\Psi
+i\Delta_4\bar{\Psi}\Psi-\Delta_5\bar{\Psi}\gamma_5\Psi.
 \end{equation}
Defining a real mass amplitude $m_0$ and angle $\theta_0$ by $\Delta_4=m_0\cos\theta_0$, $\Delta_5=m_0\sin\theta_0$, we have
\begin{align}\label{LMF}
\mathcal{L}_\textrm{MF}=i\bar{\Psi}\gamma_\mu\partial_\mu\Psi
+im_0\bar{\Psi} e^{i\theta_0\gamma_5}\Psi.
\end{align}
As mentioned in Sec.~\ref{sec:IntChiralSymm}, the pair of gamma matrices $(\Gamma_4,\Gamma_5)$ transforms as a vector under $U(1)$ chiral symmetry. Equivalently, a $U(1)$ chiral transformation can naively be compensated by a shift of $\theta_0$. Therefore, the mean-field Lagrangian (\ref{LMF}) with a fixed value of $\theta_0$ describes a state with spontaneously broken chiral symmetry. Discrete symmetries are also spontaneously broken if a gamma matrix appearing in the mean-field Hamiltonian is odd under that symmetry. A nonzero value of $\Delta_4$ breaks the antiunitary $\mathcal{T}$ and particle-hole $\mathcal{C}$ symmetries, while a nonzero value of $\Delta_5$ breaks the parity $\mathcal{P}$ symmetry. Due to our choice of gamma matrices, here the normal mass $\bar{\Psi}\Psi$ is $\mathcal{T}$-breaking while the axial mass $\bar{\Psi}\gamma_5\Psi$ is $\mathcal{T}$-preserving.

In terms of the microscopic { fermions}, the mean-field Hamiltonian (\ref{HMF}) describes a SDW ground state which spontaneously breaks translation symmetry but preserves the $SO(2)$ spin-orbit rotation symmetry. Indeed, the magnetization of the microscopic { fermions} in the $z$ direction is given by
\begin{align}
M_z(\b{r})&=\langle c_{\b{r}\alpha}^\dag\sigma_3^{\alpha\beta}c_{\b{r}\beta}\rangle
\nonumber\\
&=\langle\Psi^\dag\Gamma_4\Psi\rangle\cos 2Qz
+\langle\Psi^\dag\Gamma_5\Psi\rangle\sin 2Qz,
\end{align}
which describes a spatial modulation at wave vector $2Q$ in the direction $z$ that joins the two Weyl points. Since $\langle\Psi^\dag\Gamma_4\Psi\rangle\propto\Delta_4
=m_0\cos\theta_0$ and $\langle\Psi^\dag\Gamma_5\Psi\rangle\propto\Delta_5
=m_0\sin\theta_0$, we have
\begin{align}
M_z(\b{r})\propto m_0\cos(2Qz-\theta_0),
\end{align}
i.e., the angle $\theta_0$ corresponds physically to the phase of the SDW. Fluctuations above the mean-field ground state are described by a Lagrangian of the same form as (\ref{LMF}),
\begin{align}\label{LagFluct}
\mathcal{L}=i\bar{\Psi}\gamma_\mu\partial_\mu\Psi+im\bar{\Psi} e^{i\theta\gamma_5}\Psi,
\end{align}
but where $m$ and $\theta$ are dynamical fields. If we expand about the ground state $m(\b{r},\tau)=m_0+\delta m(\b{r},\tau)$ and $\theta(\b{r},\tau)=\theta_0+\delta\theta(\b{r},\tau)$ with $\delta m\ll m_0$ and $\delta\theta\ll 2\pi$, the amplitude fluctuations $\delta m$ are gapped and can be integrated out, while the angle fluctuations $\delta\theta$ are gapless. Indeed, $\delta\theta$ is the Goldstone mode associated with the spontaneous breaking of the $U(1)$ chiral symmetry. At energies below the single-particle gap $|m_0|$, the Goldstone mode is governed by the Lagrangian
\begin{align}
\mathcal{L}(\delta\theta)=\frac{\kappa}{2}(\partial_\mu\delta\theta)^2,
\end{align}
where the phase stiffness $\kappa$ depends on $m_0$. Fluctuations of $\delta\theta$ thus correspond physically to fluctuations of the SDW phase, i.e., the sliding mode. Sufficiently strong phase fluctuations $\delta\theta\sim 2\pi$ will melt the SDW and restore the translationally invariant Weyl semimetal ground state.

It is known that massless Dirac fermions in $2+1$ dimensions, such as those in graphene, can undergo a continuous semimetal-insulator transition similar to the one discussed here, where a mass term is spontaneously generated for sufficiently strong interactions.\cite{vafek2013} The critical point for that transition is in the universality class of the Gross-Neveu theory\cite{gross1974} in $2+1$ dimensions. The critical exponents for this strongly coupled critical point differ from those of the Gaussian fixed point and can be calculated perturbatively using the $1/N$ expansion\cite{herbut2006} or the $\epsilon$ expansion\cite{herbut2009b} where $\epsilon=1$ corresponds to $2+1$ dimensions. In $3+1$ dimensions, $\epsilon=0$ and anomalous dimensions vanish. Therefore the critical point for a continuous transition between the Weyl semimetal and the SDW state will have Gaussian critical exponents, possibly with logarithmic corrections to scaling.

The spontaneous breaking of chiral symmetry in a Weyl semimetal induced by sufficiently strong four-fermion interactions was also studied by Wang and Zhang.\cite{wang2013} In their work, a particular type of four-fermion interaction was selected that produced a CDW ground state when treated at the mean-field level. As discussed in Sec.~\ref{sec:intro}, the main difference between their approach and ours is that we consider all possible short-range interactions allowed by symmetry, and allow quantum fluctuations to determine what type of order { can} develop at strong coupling.

\subsection{Axion electrodynamics}
\label{sec:AxionEM}

  
{ Let us now assume that the fermions couple to an external electromagnetic field with charge $e$. This coupling can be reintroduced in the theory} by replacing the partial derivative $\partial_\mu$ in Eq.~(\ref{LagFluct}) by the gauge-covariant derivative $D_\mu=\partial_\mu+ieA_\mu$ where $A_0$ is the scalar potential and $A_i$ is the magnetic vector potential. This is valid in the regime, to which we restrict ourselves, where the external electromagnetic field varies slowly on the scale of the SDW wavelength $\lambda_\textrm{SDW}=\frac{\pi}{Q}$. In this limit, the Euclidean Lagrangian complete with electromagnetic fields is\cite{ChiralAnomaly}
  \begin{align}\label{LagEM}
\mathcal{L}=i\bar{\Psi}\gamma_\mu D_\mu\Psi+im\bar{\Psi} e^{i\theta\gamma_5}\Psi+\frac{1}{8\pi}(\b{E}^2+\b{B}^2),
 \end{align}
 where $\b{E}$ and $\b{B}$ are the electric and magnetic fields, respectively. As mentioned earlier, one would naively expect that the angle $\theta$ could be eliminated from the Lagrangian by a $U(1)$ chiral symmetry transformation $\Psi\rightarrow e^{-i\theta\gamma_5/2}\Psi$, $\bar{\Psi}\rightarrow\bar{\Psi}e^{-i\theta\gamma_5/2}$. However, in a quantum theory, one also has to worry about whether the integration measure $\mathcal{D}\bar{\Psi}\mathcal{D}\Psi$ in the path integral definition of the partition function remains invariant under this transformation -- if not, there is an anomaly.\cite{fujikawa1979} This is indeed what happens in our case, and the resulting anomaly is known as the chiral or Adler-Bell-Jackiw anomaly.\cite{adler1969,bell1969} The Jacobian associated with the chiral symmetry transformation gives rise to a $\b{E}\cdot\b{B}$ term in the transformed Lagrangian,\cite{kikuchi1992} and we obtain
  \begin{equation}
\mathcal{L}=i\bar{\Psi}\gamma_\mu D_\mu\Psi+im\bar{\Psi}\Psi+\frac{1}{8\pi}(\b{E}^2+\b{B}^2)
+\frac{i\theta e^2}{4\pi^2}\b{E}\cdot\b{B},
 \end{equation}
i.e., a massive Dirac fermion coupled to axion electrodynamics.\cite{wilczek1987} Strictly speaking, this derivation\cite{kikuchi1992} only holds for a spacetime-independent $\theta$, i.e., in the ground state $\theta=\theta_0$, but a perturbative calculation for a dynamical $\theta$ angle that varies slowly on the scale set by the inverse of the fermion mass $m$ gives a coupling between $\theta$ and the electromagnetic fields that has the same form.\cite{callan1985} (The anomaly calculation can also be extended to a spacetime-dependent $\theta$: see, e.g., Ref.~\onlinecite{zyuzin2012}.)

The emergence of axion electrodynamics in our effective Lagrangian points to a connection to time-reversal invariant 3D topological insulators.\cite{hasan2010,qi2011} To establish this connection, given our choice of gamma matrices it is convenient to use the chiral anomaly to rotate the mass angle by $\theta+\frac{\pi}{2}$ instead of rotating it by $\theta$. This eliminates the normal mass rather than eliminating the axial mass, and generates an axion angle of $\theta+\frac{\pi}{2}$,
\begin{align}\label{LagEMtheta}
\mathcal{L}&=i\bar{\Psi}\gamma_\mu D_\mu\Psi+m\bar{\Psi}\gamma_5\Psi+\frac{1}{8\pi}(\b{E}^2+\b{B}^2)\nonumber\\
&\hspace{4mm}
+\frac{i(\theta+\frac{\pi}{2})e^2}{4\pi^2}\b{E}\cdot\b{B}.
\end{align}
For $\theta=-\frac{\pi}{2}$, assuming $m>0$ the original Lagrangian (\ref{LagEM}) has a positive axial mass term $m\bar{\Psi}\gamma_5\Psi$, while for $\theta=\frac{\pi}{2}$, the axial mass is negative $-m\bar{\Psi}\gamma_5\Psi$. In the transformed Lagrangian (\ref{LagEMtheta}), $\theta=-\frac{\pi}{2}$ corresponds indeed to a positive axial mass with no $\b{E}\cdot\b{B}$ term, while $\theta=\frac{\pi}{2}$ corresponds to a positive axial mass but with an $\b{E}\cdot\b{B}$ term with axion angle $\theta+\frac{\pi}{2}=\pi$. Since only the axial mass is $\mathcal{T}$-preserving, a ground state with $\theta_0=-\frac{\pi}{2}$ corresponds to a $\mathcal{T}$-invariant trivial insulator, while $\theta_0=\frac{\pi}{2}$ corresponds to a $\mathcal{T}$-invariant topological insulator.\cite{qi2008,essin2009} This particular choice of topological versus trivial assumes that the vacuum outside the material can be adiabatically connected to an insulator with $\theta_0=-\frac{\pi}{2}$. Since the value of $\theta_0$ is picked by spontaneous symmetry breaking in the infinite system, all values of $\theta_0$ correspond to degenerate ground states. If we restrict ourselves to $\mathcal{T}$-invariant ground states, because the topological insulator has surface states whereas the trivial insulator does not, it seems likely that the bulk degeneracy between the two phases will be lifted by surface effects. We leave further investigation of surface effects for future work. 

All values of $\theta_0$ modulo $2\pi$ besides $\theta_0=\pm\frac{\pi}{2}$ correspond to a $\mathcal{T}$-breaking insulator.\cite{essin2010,malashevich2010} In all cases, the microscopic time-reversal symmetry is broken. Moreover, from the point of view of spontaneous chiral symmetry breaking there is nothing special about the values $\theta_0=\pm\frac{\pi}{2}$, and the ground state will generically break the $\mathcal{T}$ symmetry. The total axion angle $\theta_0+\frac{\pi}{2}$ can in principle be measured by magnetooptical Kerr and Faraday rotation.\cite{maciejko2010,tse2010}

The angle $\theta=\theta_0+\delta\theta$ is a dynamical field and the fluctuations $\delta\theta$ can be regarded as a dynamical axion field\cite{li2010,wang2011} governed by the Lagrangian
\begin{align}
\mathcal{L}(\delta\theta)=\frac{\kappa}{2}(\partial_\mu\delta\theta)^2+\frac{i(\theta_0+\frac{\pi}{2}+\delta\theta)e^2}{4\pi^2}\b{E}\cdot\b{B},
\end{align}
which was also obtained in the CDW state found in Ref.~\onlinecite{wang2013} (including the extra term $\propto 2Qz\b{E}\cdot\b{B}$ that we have omitted\cite{ChiralAnomaly}). Besides small fluctuations $\delta\ll 2\pi$, there will also be singular 1D vortex lines around which $\theta$ winds by $2\pi$. As discussed in Ref.~\onlinecite{wang2013}, these vortex lines correspond to dislocations in the SDW and are equivalent to the axion strings of particle physics\cite{witten1985} due to their direct coupling to $\b{E}\cdot\b{B}$. As discovered by Callan and Harvey,\cite{callan1985} such axion strings will trap chiral fermion modes, which could carry dissipationless current.

\section{Instabilities of the maximally chiral symmetric Hamiltonian}
\label{sec:maximal}

The noninteracting Hamiltonian (\ref{H0Dirac}) also exhibits additional chiral symmetries (Sec. \ref{sec: additional}). These additional symmetries are not respected by typical lattice-scale interactions, and thus imposing them on the interaction Hamiltonian is likely to yield a poor approximation to the true physics. Nonetheless, it is interesting to consider constraining the interaction Hamiltonian by imposing on it these additional symmetries, to see how they further reduce the number of independent couplings, and to study the instabilities of the resulting maximally chiral symmetric Hamiltonian.

\subsection{Reduction of number of independent interaction parameters by additional chiral symmetries}

Under the discrete chiral antiunitary $\tilde{\mathcal{T}}$ and particle-hole $\tilde{\mathcal{C}}$ symmetries defined in Sec.~\ref{sec: additional}, the gamma matrices transform as before, $\tilde{T}\Gamma_A\tilde{T}^{-1}=\pm\Gamma_A^*$ and $\tilde{C}\Gamma_A\tilde{C}^{-1}=\pm\Gamma_A^T$. Because the interaction terms are already constrained to be of the form $(\Psi^\dag\Gamma_A\Psi)^2$, these additional discrete chiral symmetries do not constrain the allowed couplings any further.

We now consider the chiral $SO(2)_{R-L}$ rotation symmetry. Its action on the Weyl fermions is
\begin{align}
\tilde{\mathcal{R}}(\theta)\left(\begin{array}{c}
\psi_R(\b{r}) \\
\psi_L(\b{r})
\end{array}\right)\tilde{\mathcal{R}}(\theta)^{-1}
=\tilde{R}(\theta)\left(\begin{array}{c}
\psi_R(R_\theta\b{r}) \\
\psi_L(R_{-\theta}\b{r})
\end{array}\right),
\end{align}
where $R_\theta$ is the $3\times 3$ spatial rotation matrix (\ref{SO2RotationMatrix}), and
\begin{align}
\tilde{R}(\theta)=e^{-i\theta\Gamma_3/2}.
\end{align}
Because it is a rotation in both spin space and real space, the chiral rotation symmetry only allows quartic terms with four Weyl fermions of the same chirality. This means that only the eight block-diagonal gamma matrices $\Gamma_0,\Gamma_1,\Gamma_2,\Gamma_3,\Gamma_{12},\Gamma_{13},\Gamma_{23},\Gamma_{45}$ are allowed. This reduces the number of independent couplings from nine (before using Fierz identities) to six. Expanding the quartic terms $(\Psi^\dag\Gamma_A\Psi)^2$ in Weyl components, in order to eliminate the forbidden terms of the form $\psi^\dag_R\psi_R\psi^\dag_L\psi_L$ we need to impose the three constraints $\lambda_1=\lambda_5$, $\lambda_4=\lambda_6$, and $\lambda_8=\lambda_9$ in Eq.~(\ref{VbeforeFierz}). This reduces the number of independent couplings from six to three. The three corresponding quartic terms are $X_1+X_5$, $X_4+X_6$, and $X_8+X_9$. Using Eq.~(\ref{GaussElim}), we find
\begin{align}
X_4+X_6&=-(X_1+X_5),\nonumber\\
X_8+X_9&=-2(X_1+X_5),
\end{align}
therefore after using the Fierz identities we are left with a single independent coupling corresponding to $X_1+X_5$. The Lagrangian is therefore
\begin{align}\label{eq: maximally symmetric}
\mathcal{L}=i\bar{\Psi}\gamma_\mu\partial_\mu\Psi
+g[(\bar{\Psi}\Gamma_4\Psi)^2-(\bar{\Psi}\Gamma_5\Psi)^2].
\end{align}

We now discuss the instabilities of the maximally chiral symmetric Hamiltonian (\ref{eq: maximally symmetric}). The RG equation for $g$ can be easily read off from the previous RG equations by setting $g_1=-g_2=g$ and $g_3=g_4=0$. We find that the $\mathcal{O}(g^2)$ contribution to the RG beta function vanishes. The one-loop calculation is insufficient in this case and one would need to go to higher loops. A likely possibility is that there is a flow to strong coupling $g\rightarrow\pm\infty$ for $\ell\rightarrow\infty$. One can then repeat the susceptibility analysis for this case.

\subsection{Susceptibility analysis}

For a theory governed by the one-parameter Hamiltonian (\ref{eq: maximally symmetric}), a test vertex $\Delta$ introduced in either the particle-hole or particle-particle channels renormalizes according to the flow equation
\begin{equation}
\frac{d \ln \Delta }{d\ell} = 1 + A g, 
\end{equation}
where $A$ is a numerical coefficient that depends on the channel under consideration. The appropriate value of $A$ can be obtained from the previous analysis by setting $g_1=-g_2=g$ and $g_3=g_4=0$. The strongest instability occurs in the channel with the largest positive value of $1 + A g$. 

\subsubsection{Large positive $g$}
For large positive $g$, the leading instability is in a sixfold degenerate particle-hole channel with order parameter structure $\tau_0 \otimes \bsigma$ or $\tau_3 \otimes \bsigma$. These ferromagnetic order parameters do not gap out the Weyl points but shift their position in momentum space, such that the single-particle spectrum remains gapless. Order parameters with structure $\tau_0\otimes\sigma_1=\Gamma_1$, $\tau_0\otimes\sigma_2=\Gamma_2$, and $\tau_3\otimes\sigma_3=\Gamma_3$ shift both Weyl points by the same amount in the $x$, $y$, and $z$ directions respectively, and their fluctuations are analogous to a fluctuating vector potential. This is the 3D analog of in-plane ferromagnetic order in a 2D Dirac fermion system, where ferromagnetic fluctuations also couple like a vector potential.\cite{xu2010} The resulting state breaks all three discrete $\mathcal{P},\mathcal{T},\mathcal{C}$ symmetries, and order in the $\Gamma_1$ and $\Gamma_2$ channels breaks the $SO(2)$ rotation symmetry as well due to a shift of the Weyl points away from $p_x=p_y=0$. Order parameters with structure $\tau_3\otimes\sigma_1=\Gamma_{23}$, $\tau_3\otimes\sigma_2=-\Gamma_{13}$, and $\tau_0\otimes\sigma_3=\Gamma_{12}$ shift the Weyl points relative to each other in the $x$, $y$, and $z$ directions respectively, and their fluctuations are analogous to a fluctuating chiral vector potential that couples with opposite charge to Weyl fermions of opposite chirality. The emergence of a dynamical chiral vector potential coupled to Weyl fermions was also found in a proposal to realize a Weyl semimetal in a magnetically doped topological insulator.\cite{liu2013} The resulting state preserves $\mathcal{P}$ but breaks $\mathcal{T}$ and $\mathcal{C}$, and order in the $\Gamma_{13}$ and $\Gamma_{23}$ channels also breaks the $SO(2)$ rotation symmetry due to a shift of the Weyl points away from $p_x=p_y=0$.

\subsubsection{Large negative $g$}
 If the flow is towards strong attractive coupling $g\rightarrow-\infty$, the leading instability will be in a doubly degenerate particle-particle channel, with order parameters $\langle \Psi^T \Gamma_2 \Psi \rangle$ and $\langle \Psi^T \Gamma_{13} \Psi \rangle$ that both represent spin-singlet FFLO pairing (see Appendix \ref{app:SymmSCOP}). These FFLO states were discussed previously as possible superconducting states of doped Weyl semimetals.\cite{cho2012} The bulk Bogoliubov quasiparticle spectrum obtained by diagonalizing Eq.~(\ref{HBdG}) is fully gapped in either case, $E_\pm(\b{p})=\pm\sqrt{\b{p}^2+|\Delta|^2}$ for either $\Delta_2=\Delta,\Delta_{13}=0$ or $\Delta_2=0,\Delta_{13}=\Delta$. However, the pairing amplitude in the $\Gamma_2$ channel has the same phase on each Weyl point, while it has a relative phase of $\pi$ on the two Weyl points in the $\Gamma_{13}$ channel. In the weak pairing limit, this means that at the level of the effective theory for the slow Weyl fermions, pairing in the $\Gamma_2$ channel corresponds to a trivial superconductor while pairing in the $\Gamma_{13}$ channel corresponds to a topological superconductor.\cite{qi2010} Invariance under $\mathcal{T}$ (and $\mathcal{C}$) is achieved if the pairing amplitude is pure imaginary $\Delta=i|\Delta|$ (see Appendix~\ref{app:SymmSCOP}), which corresponds to the usual spin-singlet pairing on each node $\langle\psi_R^Ti\sigma_2\psi_R\rangle
=\pm\langle\psi_L^Ti\sigma_2\psi_L\rangle\neq 0$. Either superconducting state preserves the $SO(2)$ rotation symmetry. Furthermore, $\Gamma_2$ pairing breaks the $\mathcal{P}$ symmetry while $\Gamma_{13}$ pairing preserves $\mathcal{P}$.
 
The degeneracy of the $\Gamma_2$ and $\Gamma_{13}$ channels originates from the $U(1)$ chiral symmetry of the normal state Hamiltonian (see Appendix~\ref{app:SymmSCOP}), and a given superposition of these order parameters will break this chiral symmetry spontaneously. The energetics of possible superpositions can be explored by constructing a Landau theory. 

\subsection{Landau-Ginzburg analysis}

The Landau Lagrangian, after performing a Hubbard-Stratonovich transformation and integrating out the fermions, takes the form
\begin{align}
\mathcal{L} &= - \Tr \ln \left(i\omega + H_\textrm{BdG}(\b{p})\right) + \frac{1}{4\lambda} \Tr |\Delta_2 \Gamma_2 + \Delta_{13} \Gamma_{13}|^2\nonumber\\
 &= - \Tr \ln  \left(\begin{array}{cc}i \omega + \b{p}\cdot\boldsymbol{\Gamma} & \Delta_2 \Gamma_2 + \Delta_{13} \Gamma_{13} \\ \Delta_2^* \Gamma_2 + \Delta_{13}^* \Gamma_{13} & i \omega + \b{p}\cdot\boldsymbol{\Gamma}^T \end{array} \right ) \nonumber\\ &\hspace{4mm}+ \frac{1}{4\lambda} \Tr \left(|\Delta_2|^2 + |\Delta_{13}|^2\right),
\end{align}
where $\lambda$ is the attractive coupling in the superconducting channel, and the trace is taken over spin-valley space and also represents integration over frequencies $\omega$ and momenta $\b{p}$. Working near the critical temperature $T_c$ allows us to invoke critical slowing down, and thus to set $\omega = 0$ and to integrate over momenta only. We then obtain the Landau free energy 
\begin{align}
F& = - \Tr \ln H_\textrm{BdG}(\b{p}) + \frac{1}{4\lambda} \Tr \left(|\Delta_2|^2 + |\Delta_{13}|^2\right)\nonumber\\
&=\frac{1}{4\lambda}\left(|\Delta_2|^2 + |\Delta_{13}|^2\right)
- \int\frac{d^3p}{(2\pi)^3}\nonumber\\
&\hspace{4mm}\times\ln \left[(\b{p}^2 + |\Delta_2|^2+ |\Delta_{13}|^2)^2 - (\Delta_2^*\Delta_{13}+\mathrm{c.c.})^2  \right],
\end{align}
where we used the identity $\Tr\ln M=\ln\det M$. Expanding the free energy in powers of the order parameter, we obtain 
\begin{equation}\label{FreeEnergySC}
F = F_0 \left(|\Delta_2|^2 + |\Delta_{13}|^2\right)+ K (\Delta_2^*\Delta_{13}+\mathrm{c.c.})^2,
\end{equation}
where $K$ is a strictly positive coefficient. Minimizing Eq.~(\ref{FreeEnergySC}) tells us that we should take $\Delta_2 = 0$ and $\Delta_{13}\neq 0$ (corresponding to a topological superconductor), or $\Delta_{13} = 0$ and $\Delta_2\neq 0$ (corresponding to a trivial superconductor), or we can take both order parameters to be nonzero if we give them a relative phase of $\pm\frac{\pi}{2}$. This is consistent with Eq.~(\ref{BrokenChiPairing}), and gives a fully gapped BdG spectrum $E_\pm(\b{p})=\pm\sqrt{\b{p}^2+|\Delta_2|^2+|\Delta_{13}|^2}$. Consider for example a pairing term of the form $|\Delta_2|\Psi^Ti\Gamma_2\Psi\pm i|\Delta_{13}|\Psi^Ti\Gamma_{13}\Psi+\mathrm{h.c.}$. In Weyl components, this can be written as
\begin{align}
\Delta\left(e^{i\theta_R}\psi_R^Ti\sigma_2\psi_R
+e^{i\theta_L}\psi_L^Ti\sigma_2\psi_L\right)+\mathrm{h.c.},
\end{align}
where $\Delta=\sqrt{|\Delta_2|^2+|\Delta_{13}|^2}$ and
\begin{align}
\theta_L-\theta_R=2\tan^{-1}\left(\frac{|\Delta_{13}|}{|\Delta_2|}\right),
\end{align}
implying that $\theta_L-\theta_R$ is zero for the trivial superconductor, $\pi$ for the topological superconductor, and any value in between (modulo $2\pi$) for a general superconducting state with $\Delta_2,\Delta_{13}$ both nonzero, which breaks the antiunitary $\mathcal{T}$ symmetry (note that with $|\Delta_2|=0$ and $|\Delta_{13}|\neq 0$, the apparently ``$\mathcal{T}$-odd'' pairing $\pm i|\Delta_{13}|\Psi^Ti\Gamma_{13}\Psi=\mp|\Delta_{13}|\Psi^T\Gamma_{13}\Psi$ can be made $\mathcal{T}$-even by a uniform gauge transformation). According to recent work,\cite{qi2013} the electromagnetic response of this fully gapped $\mathcal{T}$-breaking superconductor should be of the ``Higgs-axion'' type,
\begin{align}
\mathcal{L}_\textrm{eff}&=\frac{1}{2}\rho_R(\partial_\mu\theta_R-2eA_\mu)^2+\frac{1}{2}\rho_L(\partial_\mu\theta_L-2eA_\mu)^2\nonumber\\
&\hspace{4mm}+\frac{i(\theta_L-\theta_R)e^2}{8\pi^2}\b{E}\cdot\b{B}.
\end{align}
As in Sec.~\ref{sec:AxionEM}, however, all superconducting states considered here break the microscopic time-reversal symmetry. We speculate that this is manifest in the electromagnetic response in the presence of an additional $\propto 2Qz\b{E}\cdot\b{B}$ term coming from the FFLO nature of the pairing that is missed in our description in terms of slow fields $\psi_R,\psi_L$.\cite{ChiralAnomaly} We leave this question, as well as the analysis of fluctuation corrections to the free energy that will probably lift the degeneracy between the various scenarios, for future work.

\section{Conclusion}

We have presented a low-energy description of interacting { fermions} in the simplest type of time-reversal symmetry-breaking Weyl semimetal with two Weyl points related by inversion symmetry. We restricted ourselves to short-range interactions and used the symmetries of the noninteracting low-energy Hamiltonian, which consisted of two continuum Weyl fermions of opposite chirality, to constrain the form of the interaction term. Combined with the use of Fierz identities, this reduced the number of independent coupling constants from 136 to four, which made the problem amenable to an analytical perturbative RG analysis. We computed the RG beta functions to one loop, and found a single stable trajectory in the four-dimensional coupling constant space towards strong coupling. We computed the susceptibilities for all possible momentum-independent order parameters in the particle-hole and particle-particle channels, and determined that the leading instability was towards SDW ordering. Using an anomaly calculation, we found that the sliding mode or SDW phase mode $\theta$ coupled to external electromagnetic fields via an axion term $\propto\theta\b{E}\cdot\b{B}$ similar to that found previously by Wang and Zhang\cite{wang2013} for CDW order. 

We also investigated a maximally chiral symmetric model with a single independent coupling constant. While ``true'' lattice scale interactions are unlikely to display maximal chiral symmetry, this simplified problem nevertheless presents an interesting toy model for theoretical study. In this case, the susceptibility analysis indicated two possible types of ground states: gapless ferromagnetic states, or a gapped superconducting state. The spin waves of the gapless ferromagnetic states were found to couple to the fermionic quasiparticles like the spatial components of a normal or chiral gauge field,\cite{liu2013} depending on the type of ferromagnetic order. The superconducting state was found to be of the exotic FFLO type, with a ``Higgs-axion'' type electromagnetic response. 

\begin{acknowledgments}
We thank V. Aji and O. Vafek for useful discussions. This work was supported by the Simons Foundation (JM) and by a PCTS fellowship (RN).
\end{acknowledgments}

\appendix
\section{Gamma matrix identities}
\label{app:gamma}

For convenience, we reproduce here certain gamma matrix identities from Appendix A of Ref.~\onlinecite{murakami2004} that are used extensively throughout the paper:
\begin{align}
[\Gamma_{ab},\Gamma_c]&=2i(\delta_{ac}\Gamma_b-\delta_{bc}\Gamma_a),\label{EqGammaA1}\\
\{\Gamma_{ab},\Gamma_c\}&=\epsilon_{abcde}\Gamma_{de},\\
[\Gamma_{ab},\Gamma_{cd}]&=-2i(\delta_{bc}\Gamma_{ad}
-\delta_{bd}\Gamma_{ac}\nonumber\\
&\hspace{4mm}-\delta_{ac}\Gamma_{bd}+\delta_{ad}\Gamma_{bc}),
\label{EqGammaComm}
\\
\{\Gamma_{ab},\Gamma_{cd}\}&=2\epsilon_{abcde}\Gamma_e
+2\delta_{ac}\delta_{bd}-2\delta_{ad}\delta_{bc},\\
\Tr(\Gamma_a\Gamma_b)&=4\delta_{ab},\\
\Tr(\Gamma_a\Gamma_b\Gamma_c)&=0,\\
\Tr(\Gamma_a\Gamma_b\Gamma_c\Gamma_d)&=
4(\delta_{ab}\delta_{cd}+\delta_{ad}\delta_{bc}
-\delta_{ac}\delta_{bd}),\\
\Tr(\Gamma_a\Gamma_b\Gamma_c\Gamma_d\Gamma_e)&=
-4\epsilon_{abcde},
\end{align}
where $\epsilon_{abcde}$ is the totally antisymmetric symbol in five dimensions and $\epsilon_{12345}=+1$.

\section{Symmetries of the superconducting order parameters}
\label{app:SymmSCOP}

In this Appendix we discuss how superconducting order parameters transform under the symmetries of the Hamiltonian. We restrict ourselves to superconducting order parameters of the form $\langle\Psi^T\Gamma_A\Psi\rangle$ that are momentum-independent in the low-energy effective theory, corresponding to on-site pairing. This does not necessarily mean that the resulting superconducting state is conventional. Matrices $\Gamma_A$ with nonzero diagonal blocks describe pairing between { fermions} on the same Weyl point, which corresponds to an exotic FFLO state\cite{fulde1964,larkin1965} that spontaneously breaks translation symmetry in the $z$ direction, due to the fact that Cooper pairs carry nonzero center-of-mass momentum in this direction. Indeed, pairing between the slow Weyl fermion operators $\psi_R,\psi_L$ translates into pairing between the microscopic { fermion} operators $c$ that is given by
\begin{align}
\langle c_{\b{r}\alpha}c_{\b{r}\beta}\rangle&\simeq
e^{2iQz}\langle\psi_{R\alpha}\psi_{R\beta}\rangle
+e^{-2iQz}\langle\psi_{L\alpha}\psi_{L\beta}\rangle\nonumber\\
&\hspace{4mm}+\langle\psi_{R\alpha}\psi_{L\beta}\rangle
+\langle\psi_{L\alpha}\psi_{R\beta}\rangle,
\end{align}
where we have used Eq.~(\ref{ExpandElectron}).

The mean-field Hamiltonian for a superconducting state is
\begin{align}
H_\textrm{MF}=H_0+\frac{1}{2}\int d^3r\,(\Psi^T\Delta^\dag\Psi+\textrm{h.c.}),
\end{align}
where $\Delta=-\Delta^T$ by Fermi statistics, but is otherwise arbitrary. A general $4\times 4$ antisymmetric matrix can be expanded as $\Delta=\Delta_A\Gamma_A$, where $\Gamma_A$ are linearly independent antisymmetric $4\times 4$ Hermitian matrices, and $\Delta_A$ is in general complex. Among the sixteen linearly independent $4\times 4$ Hermitian matrices, six of them are antisymmetric: $\Gamma_2,\Gamma_5,\Gamma_{13},\Gamma_{14},\Gamma_{25},\Gamma_{34}$. In terms of these antisymmetric $\Gamma$ matrices, the momentum space Hamiltonian reads
\begin{align}
H_\textrm{MF}=\frac{1}{2}\int\frac{d^3p}{(2\pi)^3}\Phi^\dag(\b{p})
H_\textrm{BdG}(\b{p})\Phi(\b{p}),
\end{align}
where $\Phi(\b{p})$ is a 8-component Nambu spinor defined as
\begin{align}
\Phi(\b{p})=\left(\begin{array}{c}
\Psi(\b{p}) \\
(\Psi^\dag(-\b{p}))^T
\end{array}\right),\,
\Phi^\dag(\b{p})=\left(\begin{array}{cc}
\Psi^\dag(\b{p}) & \Psi^T(-\b{p})
\end{array}\right),
\end{align}
and the $8\times 8$ Bogoliubov-de Gennes (BdG) Hamiltonian matrix is
\begin{align}\label{HBdG}
H_\textrm{BdG}(\b{p})=\left(\begin{array}{cc}
\b{p}\cdot\boldsymbol{\Gamma} & \sum_A \Delta_A\Gamma_A \\
\sum_A \Delta_A^*\Gamma_A & \b{p}\cdot\boldsymbol{\Gamma}^T
\end{array}\right),
\end{align}
where $\b{p}\cdot\boldsymbol{\Gamma}=p_x\Gamma_1
+p_y\Gamma_2+p_z\Gamma_3$.

We wish to determine how the superconducting order parameter transforms under the symmetries of our problem. We first consider the rotation symmetry of Sec.~\ref{sec:RotSymm}, under which a typical pairing term transforms as
\begin{align}
&\mathcal{R}(\theta)\int d^3r\left(\Delta_A^*\Psi^T\Gamma_A\Psi+\mathrm{h.c.}\right)
\mathcal{R}(\theta)^{-1}\nonumber\\
&=\int d^3r\left(\Delta_A^*\Psi^T(R_\theta\b{r})R(\theta)^T\Gamma_AR(\theta)
\Psi(R_\theta\b{r})+\mathrm{h.c.}\right)\nonumber\\
&=\int d^3r'\left(\Delta_A^*\Psi^T(\b{r}')R(\theta)^T\Gamma_AR(\theta)
\Psi(\b{r}')+\mathrm{h.c.}\right),
\end{align}
where $\b{r}'=R_\theta\b{r}$. Since $R(\theta)=e^{-i\theta\Gamma_{12}/2}$ [Eq.~(\ref{Rtheta})] and $\Gamma_{12}=\Gamma_{12}^T$, we have
\begin{align}\label{RTGAR}
R(\theta)^T\Gamma_AR(\theta)&=e^{-i\theta\Gamma_{12}/2}
\Gamma_A e^{-i\theta\Gamma_{12}/2}\nonumber\\
&=\cos^2(\theta/2)\Gamma_A-\sin^2(\theta/2)
\Gamma_{12}\Gamma_A\Gamma_{12}\nonumber\\
&\hspace{4mm}-i\sin(\theta/2)\cos(\theta/2)\{\Gamma_{12},\Gamma_A\}.
\end{align}
Using the algebra of gamma matrices, we find that $\Gamma_2,\Gamma_{13},\Gamma_{14},\Gamma_{25}$ transform as scalars,
\begin{align}
R(\theta)^T\Gamma_AR(\theta)=\Gamma_A,\hspace{5mm}
A=2,13,14,25,
\end{align}
hence the associated pairing terms preserve rotation symmetry. On the other hand, $(\Gamma_5,\Gamma_{34})$ are related by a rotation,
\begin{align}
R(\theta)^T\Gamma_5R(\theta)&=
\cos\theta\Gamma_5-i\sin\theta\Gamma_{34},\label{SCRotGam5}\\
R(\theta)^T\Gamma_{34}R(\theta)&=
\cos\theta\Gamma_{34}-i\sin\theta\Gamma_5,\label{SCRotGam34}
\end{align}
which means that if we consider a pairing term of the form
\begin{align}\label{nonzeroJzpairing}
\int d^3r\left(\Delta_5^*\Psi^T\Gamma_5\Psi\mp i\Delta_{34}^*\Psi^T\Gamma_{34}\Psi+\mathrm{h.c.}\right),
\end{align}
under a rotation this term keeps the same form but with $\Delta_5,\Delta_{34}$ replaced by $\Delta_5',\Delta_{34}'$ where
\begin{align}\label{SCOPtransformRot}
\left(\begin{array}{c}
\Delta_5' \\
\Delta_{34}'
\end{array}\right)=
\left(\begin{array}{cc}
\cos\theta & \mp\sin\theta \\
\pm\sin\theta & \cos\theta
\end{array}\right)
\left(\begin{array}{c}
\Delta_5 \\
\Delta_{34}
\end{array}\right),
\end{align}
which preserves $|\Delta_5'|^2+|\Delta_{34}'|^2=|\Delta_5|^2+|\Delta_{34}|^2$. Therefore $(\Delta_5,\Delta_{34})$ transform as a vector under rotations. 
The choice of a particular linear combination of $\Delta_5$ and $\Delta_{34}$ breaks $SO(2)$ rotation symmetry spontaneously.

A similar analysis can be done for the $U(1)$ chiral symmetry of Sec.~\ref{sec:ChiralSymm}. A typical pairing term transforms as
\begin{align}
\mathcal{R}_\chi(\phi)\Psi^T\Gamma_A\Psi
\mathcal{R}_\chi(\phi)^{-1}=\Psi^T R_\chi(\phi)^T\Gamma_AR_\chi(\phi)\Psi.
\end{align}
Since $R_\chi(\phi)=e^{-i\phi\Gamma_{45}/2}$ [Eq.~(\ref{Rchiral})] and $\Gamma_{45}=\Gamma_{45}^T$, we have similarly to Eq.~(\ref{RTGAR}),
\begin{align}
R_\chi(\phi)^T\Gamma_AR_\chi(\phi)&=e^{-i\phi\Gamma_{45}/2}
\Gamma_A e^{-i\phi\Gamma_{45}/2}\nonumber\\
&=\cos^2(\phi/2)\Gamma_A-\sin^2(\phi/2)
\Gamma_{45}\Gamma_A\Gamma_{45}\nonumber\\
&\hspace{4mm}-i\sin(\phi/2)\cos(\phi/2)\{\Gamma_{45},\Gamma_A\}.
\end{align}
We find that $\Gamma_5,\Gamma_{14},\Gamma_{25},\Gamma_{34}$ transform as scalars under chiral symmetry, while $(\Gamma_2,\Gamma_{13})$ transform as
\begin{align}
R_\chi(\phi)^T\Gamma_2R_\chi(\phi)&=
\cos\phi\Gamma_2+i\sin\phi\Gamma_{13},\\
R_\chi(\phi)^T\Gamma_{13}R_\chi(\phi)&=
\cos\phi\Gamma_{13}+i\sin\phi\Gamma_2.
\end{align}
If we consider a pairing term of the form
\begin{align}\label{BrokenChiPairing}
\int d^3r\left(\Delta_2^*\Psi^T\Gamma_2\Psi\pm i\Delta_{13}^*\Psi^T\Gamma_{13}\Psi+\mathrm{h.c.}\right),
\end{align}
under chiral symmetry $(\Delta_2,\Delta_{13})$ transform as in Eq.~(\ref{SCOPtransformRot}) but with $\theta$ replaced by $\phi$. Therefore the pairing (\ref{BrokenChiPairing}) describes a superconducting state that breaks $U(1)$ chiral symmetry spontaneously.

We now consider the discrete symmetrices $\mathcal{P},\mathcal{T},\mathcal{C}$. Under parity $\mathcal{P}$, a typical pairing term transforms as
\begin{align}
&\mathcal{P}\int d^3r\left(\Delta_A^*\Psi^T\Gamma_A\Psi+\mathrm{h.c.}\right)
\mathcal{P}^{-1}\nonumber\\
&=\int d^3r\left(\Delta_A^*\Psi^T(-\b{r})P^T\Gamma_AP\Psi(-\b{r})
+\mathrm{h.c.}\right)\nonumber\\
&=\int d^3r\left(\Delta_A^*\Psi^T(\b{r})P^T\Gamma_AP\Psi(\b{r})
+\mathrm{h.c.}\right),
\end{align}
hence the pairing is even-parity (preserves parity) if $P^T\Gamma_AP=\Gamma_A$ and odd-parity (breaks parity) if $P^T\Gamma_AP=-\Gamma_A$. Under the antiunitary symmetry $\mathcal{T}$, we have
\begin{align}
\mathcal{T}\left(\Delta_A^*\Psi^T\Gamma_A\Psi+\mathrm{h.c.}\right)
\mathcal{T}^{-1}=\Delta_A\Psi^TT^T\Gamma_A^*T\Psi
+\mathrm{h.c.},
\end{align}
hence a superconducting state is invariant under $\mathcal{T}$ if the pairing amplitude is real $\Delta_A=\Delta_A^*$ and $\Gamma_A$ is even under $T$, i.e., $T^T\Gamma_A^*T=\Gamma_A$, or if the pairing amplitude is pure imaginary $\Delta_A=-\Delta_A^*$ and $\Gamma_A$ is odd under $T$, i.e., $T^T\Gamma_A^*T=-\Gamma_A$. Under particle-hole symmetry $\mathcal{C}$, we have
\begin{align}
\mathcal{C}\left(\Delta_A^*\Psi^T\Gamma_A\Psi+\mathrm{h.c.}\right)
\mathcal{C}^{-1}=\Delta_A\Psi^TC^T\Gamma_AC\Psi
+\mathrm{h.c.},
\end{align}
thus the conditions for invariance under $\mathcal{C}$ are the same as those for invariance under $\mathcal{T}$, with $\Gamma_A$ being even under $C$ if $C^T\Gamma_AC=\Gamma$ and odd if $C^T\Gamma_AC=-\Gamma_A$. We summarize the transformation properties of the pairing matrices under the discrete symmetries in Table~\ref{tableCPT}.
\begin{table}[t]
\begin{tabular}{c||cccccc}
\hline
 & $\Gamma_2$ & $\Gamma_5$ & $\Gamma_{13}$ & $\Gamma_{14}$ & $\Gamma_{25}$ & $\Gamma_{34}$ \\
\hline\hline
$P$ & $-$ & $-$ & $+$ & $-$ & $+$ & $-$  \\ \hline
$T$ & $-$ & $+$ & $-$ & $-$ & $+$ & $-$ \\ \hline
$C$ & $-$ & $+$ & $-$ & $-$ & $+$ & $-$\\ \hline
\end{tabular}
\caption{The pairing matrices $\Gamma_A$ are either even ($+$) or odd ($-$) under parity ($P$), antiunitary symmetry ($T$), and particle-hole symmetry ($C$).}
\label{tableCPT}
\end{table}

Finally, we consider the enhanced $SO(3)$ rotation symmetry that emerges asymptotically along the stable RG flow to strong coupling and makes the $\Gamma_{14}$ pairing channel degenerate with the $\Gamma_5$ and $\Gamma_{34}$ channels. Along this flow, the Hamiltonian is given by [see Eq.~(\ref{HFixedTrajectory})]
\begin{align}\label{HFixedTrajectory2}
H=\int d^3r&\bigl[\Psi^\dag\boldsymbol{\Gamma}\cdot(-i\nabla)\Psi+g_A(\Psi^\dag\Psi)^2
+g_B(\Psi^\dag\Gamma_{45}\Psi)^2\nonumber\\
&+g_C(\Psi^\dag\boldsymbol{\Gamma}\Psi)^2\bigr],
\end{align}
where $\boldsymbol{\Gamma}=(\Gamma_1,\Gamma_2,\Gamma_3)$ and we have set $v_\parallel=v_z=1$ as in our RG analysis. We now show that this Hamiltonian commutes with the $SO(3)$ rotation operator $\mathcal{R}(\hat{\b{n}},\theta)$ defined by
\begin{align}\label{SO3trans}
\mathcal{R}(\hat{\b{n}},\theta)\Psi(\b{r})\mathcal{R}(\hat{\b{n}},\theta)^{-1}=R(\hat{\b{n}},\theta)\Psi(R_{\hat{\b{n}},\theta}\b{r}),
\end{align}
which describes a rotation by an angle $\theta\in[0,2\pi)$ around the axis specified by the unit vector $\hat{\b{n}}$. The $4\times 4$ representation matrix $R(\hat{\b{n}},\theta)$ is given by
\begin{align}
R(\hat{\b{n}},\theta)=e^{-i\theta\hat{\b{n}}\cdot\boldsymbol{\Sigma}/2},
\end{align}
where
\begin{align}
\Sigma_i=\frac{1}{2}\epsilon_{ijk}\Gamma_{jk},\hspace{5mm}
i=1,2,3.
\end{align}
We have $\Sigma_1=\Gamma_{23}$, $\Sigma_2=-\Gamma_{13}$, and $\Sigma_3=\Gamma_{12}$ that we recognize as the generator of $SO(2)$ rotations about the $z$ axis (Sec.~\ref{sec:RotSymm}). $R_{\hat{\b{n}},\theta}$ is the standard $3\times 3$ rotation matrix given for small $\theta$ by
\begin{align}
R_{\hat{\b{n}},\theta}^{ik}=\delta_{ik}+\theta\epsilon_{ijk}\hat{n}_j
+\mathcal{O}(\theta^2).
\end{align}
The $g_A$ term in Eq.~(\ref{HFixedTrajectory2}) is manifestly invariant under the unitary transformation (\ref{SO3trans}). Using Eq.~(\ref{EqGammaComm}), we have $[\Gamma_{45},\Sigma_i]=0$ and the $g_B$ term is invariant as well. Using Eq.~(\ref{EqGammaA1}), we find
\begin{align}
e^{i\theta\hat{\b{n}}\cdot\boldsymbol{\Sigma}/2}\Gamma_i
e^{-i\theta\hat{\b{n}}\cdot\boldsymbol{\Sigma}/2}
=R_{\hat{\b{n}},\theta}^{ik}\Gamma_k,
\end{align}
hence the $g_C$ term transforms as
\begin{align}
(\Psi^\dag\boldsymbol{\Gamma}\Psi)^2&\rightarrow(\Psi^{\dag\prime}
R_{\hat{\b{n}},\theta}^{ik}\Gamma_k\Psi')(\Psi^{\dag\prime}R_{\hat{\b{n}},\theta}^{i\ell}\Gamma_\ell\Psi')\nonumber\\
&=(\Psi^{\dag\prime}\Gamma_k\Psi')(\Psi^{\dag\prime}\Gamma_\ell\Psi')(R_{\hat{\b{n}},\theta}^TR_{\hat{\b{n}},\theta})_{k\ell}\nonumber\\
&=(\Psi^{\dag\prime}\boldsymbol{\Gamma}\Psi')^2,
\end{align}
since the rotation matrix $R_{\hat{\b{n}},\theta}$ is orthogonal, and the rotation of coordinates $\Psi\rightarrow\Psi'=\Psi(R_{\hat{\b{n}},\theta}\b{r})$ can be absorbed by a change of integration variables in Eq.~(\ref{HFixedTrajectory2}). Finally, the derivative operator $\nabla$ transforms oppositely to $\boldsymbol{\Gamma}$ under rotations and the kinetic term in Eq.~(\ref{HFixedTrajectory2}) is also $SO(3)$ invariant.

We have already seen that $\Gamma_5$ and $\Gamma_{34}$ transform into each other under an $SO(2)$ rotation about the $z$ axis [Eq.~(\ref{SCRotGam5})-(\ref{SCRotGam34})]. Under a rotation about the $x$ axis, $\Gamma_5$ transforms as
\begin{align}
R(\hat{\b{x}},\theta)^T\Gamma_5R(\hat{\b{x}},\theta)&=e^{-i\theta\Gamma_{23}/2}\Gamma_5e^{-i\theta\Gamma_{23}/2}\nonumber\\
&=\cos^2(\theta/2)\Gamma_5-\sin^2(\theta/2)\Gamma_{23}
\Gamma_5\Gamma_{23}\nonumber\\
&\hspace{4mm}-i\sin(\theta/2)\cos(\theta/2)\{\Gamma_{23},\Gamma_5\}\nonumber\\
&=\cos\theta\Gamma_5-i\sin\theta\Gamma_{14},
\end{align}
hence the $\Gamma_{14}$ pairing channel must be degenerate with the $\Gamma_5$ and $\Gamma_{34}$ channels.

\bibliography{weyl}

\begin{thebibliography}{84}
\bibitem [{Not()}]{NoteSC}%
  \BibitemOpen
  \href@noop {} {}\bibinfo {note} {Here we must exclude arbitrarily weak
  attractive electron-electron interactions that lead to the Cooper
  instability, as well as nested Fermi surfaces that are unstable to
  arbitrarily weak repulsive electron-electron interactions.}\BibitemShut
  {Stop}%
\end{thebibliography}

\end{document}